\documentclass[usenatbib]{aastex6}

\usepackage{natbib}

\begin{document}

\title{Ages of 70 dwarfs of three populations in the solar neighborhood: considering O and C abundances in stellar models}

\author{Z. S. Ge\altaffilmark{1}, S. L. Bi\altaffilmark{1}\thanks{E-mail:bisl@bnu.edu.cn}, Y. Q. Chen\altaffilmark{2}, T. D. Li\altaffilmark{3,4,5}, J. K. Zhao\altaffilmark{2}, K. Liu\altaffilmark{1}, Fergusion J. W.\altaffilmark{6}, Y. Q. Wu\altaffilmark{1} }

\affil{ }

\altaffiltext{1}{Department of Astronomy, Beijing Normal University, Beijing 100875, China}
\altaffiltext{2}{Key Laboratory of Optical Astronomy, National Astronomical Observatories, Chinese Academy of Sciences, Beijing 100012, China}
\altaffiltext{3}{Sydney Institute for Astronomy (SIfA), School of Physics, University of Sydney, NSW 2006, Australia}
\altaffiltext{4}{Stellar Astrophysics Centre, Department of Physics and Astronomy, Aarhus University, Ny Munkegade 120, DK-8000 Aarhus C, Denmark}
\altaffiltext{5}{Key Laboratory of Solar Activity, National Astronomical Observatories, Chinese Academy of Science, Beijing 100012, China}
\altaffiltext{6}{Department of Physics, Wichita State University, Wichita, KS 67260-0032, USA}

\begin{abstract}

Oxygen and carbon are important elements in stellar populations. Their behavior refers to the formation history of the stellar populations. C and O abundances would also obviously influence stellar opacities and the overall metal abundance $Z$. With observed high-quality spectroscopic properties, we construct stellar models with C and O elements, to give more accurate ages for 70 metal-poor dwarfs, which have been determined to be high-$\alpha$ halo, low-$\alpha$ halo and thick-disk stars. Our results show that high-$\alpha$ halo stars are relatively older than low-$\alpha$ halo stars by around 2.0 Gyr. The thick-disk population has an age range between the two halo populations. The age distribution profiles indicate that high-$\alpha$ halo and low-$\alpha$ halo stars match the in situ accretion simulation by Zolotov et al., and the thick-disk stars might be formed in a relatively quiescent and long-lasting process. We also note that stellar ages are very sensitive to O abundance, since the ages clearly increase with increasing [O/Fe] values. Additionally, we obtain several stars with peculiar ages, including 2 young thick-disk stars and 12 stars older than the universe age.

\end{abstract}


\section{Introduction} \label{sec:intro}

Knowledge of various stellar populations is of utmost importance for understanding the formation and evolution of the Galaxy. The Galactic halo preserves the information of the very early universe. Two classic works presented two types of formation histories for the Galactic halo: \citet{Eggen1962} stated that the halo was formed in the monolithic collapse model, \citet{SearleZinn1978} suggested that the halo contains a collapsed inner halo and an outer halo accreted from dwarf galaxies. The origin and structure of the Galactic disk are still debated: whether the disk are constructed by two main structures, i.e. thick disk and thin disk (e.g., \citealt{GR1983}; \citealt{J2008}), or if there is no distinct thick disk \citep{Bovy2012}. Large-scale surveys of the Galaxy, such as the LAMOST survey (\citealt{Deng2012}; \citealt{Zhao2012}; \citealt{Liu2014}), the Sloan Digital Sky Survey (SDSS) \citep{Yanny2009}, and the $Gaia$ mission (\citealt{Perryman2001}; \citealt{Lindegren2010}) help to fully describe the picture of the Galaxy. These surveys provide us high-quality observations to study the Galaxy structure and evolution by tracing the stellar populations.

Stars belonging to one population are thought to be born homologously, and are sharing similar chemical composition and dynamics. Chemical abundances, especially from those of F, G, K dwarfs, are long scaled records of the Galactic chemical evolution. The abundance ratios of oxygen and carbon, as well as $\alpha$-elements (i.e. Ne, Mg, Si, S, Ca and Ti) are indicators of the stellar birth environment and formation scenarios (e.g., \citealt{Kobayashi2006}; \citealt{Ramirez12}). The kinematical characters help us to understand the activities of the stars (\citealt{Bensby2003}). The chemistry and dynamics are important tools to distinguish different populations. Furthermore, with population ages, the properties of the stellar populations can be understood more comprehensively (e.g., \citealt{Schuster2012}; \citealt{Haywood2013}; \citealt{Ramirez13}; \citealt{Hawkins2014}).

The age of a stellar population is a key character to study the structure and evolution process of the population.
Many works suggested that the Galactic halo is composed of a complex process with accreted components after the initial collapse \citep{Ibata1994}. Using SDSS low-resolution spectra, \citet{JW2011} found some stars with bluer colors and younger age than the dominant population of halo stars, and suggested that they may be accreted from external galaxies. With high-resolution spectra, series works, i.e., \citet{NS10,NS11,NS12} and \citet{Schuster2012}, confirmed that two halo populations exist that are distinguished by element abundances, ages, and orbital parameters. Using a sample of F and G dwarfs from low-resolution SDSS spectra, \citet{Hawkins2014} found that the relatively different $\alpha$-rich and $\alpha$-poor populations in the Galactic halo have different ages in the metal-rich edge and have identical ages in the metal-poor region. Whether the Galactic disk consists of a thick disk and a thin disk that were formed in different processes is still unclear. Distinguished by stellar ages, \citet{Fuhrmann2008} have found that the old population, which is regarded as thick-disk population, has higher $\alpha$-enhancement based on the stellar ages. \citet{Haywood2013} distinguished the stellar populations by [$\alpha$/Fe] and noted that different populations have different age-[$\alpha$/Fe] profiles. \citet{Ramirez13} assigned a kinematic probability for the stars to be thin-disk, thick-disk, and halo stars and confirmed that thick-disk stars have relatively higher [O/Fe] and older ages. However, both \citet{Haywood2013} and \citet{Ramirez13} found some stars with thick-disk features but younger ages, and some stars with thin-disk features but older ages. Maybe, as suggested by \citet{Bovy2012,Bovy2016}, the Galactic disk has no distinct thick-disk stellar population.

To obtain accurate stellar ages, we need to know the evolution stage of the star by constructing stellar models with adequate input physics and chemical compositions. The metal mixture used in models influences the opacities and overall metallicity $Z$ of the models. One of the most popular metal mixture patterns is the $\alpha$-enhanced metal mix, which is generally used for the metal-poor stars in the solar neighborhood. The traditional $\alpha$-enhanced metal mix considered that oxygen has the same enrichment factor as $\alpha$-elements (e.g., $Y^2$ isochrones, \citealt{Yi2001,Yi2003}, \citealt{Kim2002}, \citealt{Demarque2004}; Dartmouth Stellar Evolution Database, \citealt{Dotter2008}; Padova stellar models, \citealt{Girardi2000}, \citealt{Salasnich2000}), because observed O abundances are not easily available. However, other models consider oxygen behaving differently from $\alpha$-elements, because C, N, and O elements contribute more to the whole metal mixture. For example, to make a more accurate estimation for the ancient dwarf halo stars, \citet{VandenBerg2014} consider that oxygen has a higher enhancement factor than the $\alpha$-elements based on observations. To explain the multiple populations in some Globular Clusters, CNO-enhanced models were adopted (e.g., \citealt{Salaris2006}; \citealt{Ventura2009}). From all these works we know that to obtain stellar ages more accurately, we need high-quality atmospheric parameters of the star, especially the chemical compositions.

With high-resolution UVES-FIES spectra, \citet{NS10} compiled a sample, hereafter the NS10 sample, that they divided into high-$\alpha$ halo and low-$\alpha$ halo stars using $\alpha$-elements (i.e., Mg, Si, Ca, and Ti), as well as other elements such as Na, Ni, Cu, Zn, and Ba \citep{NS11}. The NS10 sample also includes some thick-disk stars according to the space velocities of the stars. \citet{Schuster2012} have estimated the age values of the NS10 sample with $Y^2$ isochrones for stars within the $Y^2$ range. In addition, the very cool samples ($T_{\rm{eff}}$ $<$ 5600 K) were estimated with manual corrections. \citealp{NS12} obtained the masses of the NS10 sample after correcting for all the effective temperatures (adding 100 K to the original $T_{\rm{eff}}$). \citet{Nissen14} presented observed C and O abundances for NS10 sample and additionally included an amount of thick- and thin-disk stars from HARPS-FEROS spectra, called the Nissen14 sample.

With updated and self-consistent atmospheric parameters and C O abundances, we are able to obtain accurate fundamental parameters for the metal-poor populations, i.e., high-$\alpha$ halo, low-$\alpha$ halo, and thick-disk population, from the Nissen14 sample. We use metal mixtures including C and O abundances in stellar models, as well as $\alpha$-enhanced $Y^2$ isochrones to estimate stellar parameters. With stellar ages, we assume the probable formation scenarios and evolution process for different populations. In Section 2 we introduce the observation properties of the Nissen14 sample. In Section 3 we construct stellar models and obtain fundamental parameters of our sample stars. In Section 4 we analyze the ages of different stellar populations, explaining the formation and evolution process for each population. We present the conclusions of this work in Section 5.

\section{Observations}

\subsection{The Nissen14 sample}

\citet{Nissen14} determined C and O abundances for more than 100 F and G main-sequence stars in the solar neighborhood with -1.6 $<$ [Fe/H] $<$ +0.4 selected from the HARPS-FEROS sample and the UVES-FIES sample. The O abundances are derived from the forbidden [O \i] line at 6300 \AA and the $\lambda$7774 O \i~triplet lines. The C abundances are measured from C \i~lines at 5052 and 5380\AA. All the C and O abundances have non-local thermal equilibrium (non-LTE) corrected results.

The HARPS-FEROS stars in the Nissen14 sample consist of thick- and thin-disk populations classified by their [$\alpha$/Fe]-[Fe/H] relation \citep[their Figure 1]{Adibekyan2013}. The HARPS spectra cover a wavelength range from 3800 to 6900 \AA~with a resolution of $R$ $\simeq$ 115,000 \citep{Mayor2003}. The FEROS spectra have $R$ $\simeq$ 48,000 and a typical signal-to-noise ratios (S/N) of 200 \citep{Kaufer1999}. For these stars, photometric values based on H\i PPARCOS parallaxes are applied. The typical errors for $T_{\rm{eff}}$, log $g$, and [Fe/H] of stars from the HARPS-FEROS spectra are 30 K, 0.05 dex, and 0.03 dex, respectively.

Most stars of the UVES-FIES sample belong to the halo population, according to the high space velocities with respect to the local standard of rest ($V_{\rm{LSR}}$ $>$ 180 kms$^{-1}$), but 16 stars have thick-disk kinematics. The halo stars have been determined as high-$\alpha$ and low-$\alpha$ populations using chemical compositions, stellar ages, and orbital parameters. The VLT/UVES spectra have resolutions $R$ $\simeq$ 55,000 and an S/N from 250 to 500. The FIES spectragraph at the Nordic Optical Telescope provide a resolution of $R$ $\simeq$ 40,000 and an S/N $\simeq$ 140-200. The atmospheric properties for UVES-FIES stars are obtained through a spectroscopic method. The typical errors for T$_{\rm{eff}}$, log $g$, and [Fe/H] of stars from UVES-FIES spectra are 35 K, 0.06 dex, and 0.03 dex, respectively.

\subsection{Oxygen and Carbon Behaviors}

The metal-poor populations in the Nissen14 sample, i.e., high-$\alpha$ halo, low-$\alpha$ halo, and thick-disk population, have interesting behaviors for C and O abundances. The [C/Fe] and [O/Fe] values seem to fit a line for stars with [Fe/H] $<$ -0.8, as presented in \citet[their Figure 11]{Nissen14}. Stars of the high-$\alpha$ halo and the thick-disk population with -0.8 $<$ [Fe/H] $<$ -0.2 have [O/Fe] values that decrease with increasing [Fe/H], and their C abundances have non-negligible dispersion \citet[their Figures 9, 10]{Nissen14}.

We adopt non-LTE-corrected [C/H] and [O/H] as observation abundances. For stars with [O/H]$_{\rm{6300}}$, we use the average value from [O/H]$_{\rm{7774}}$ and [O/H]$_{\rm{6300}}$ as O abundance. The [$\alpha$/Fe] values for the UVES-FIES sample has been provided by \citet{NS10}, by making an average of [Mg/Fe], [Si/Fe], [Ca/Fe], and [Ti/Fe]. For the HARPS-FEROS sample, we use [Mg/Fe], [Si/Fe], and [Ti/Fe] from \citet{Adibekyan2012} to calculate [$\alpha$/Fe]. A few thick-disk stars with [Fe/H] $>$ -0.2 dex were deleted, because in the metal-rich region the thick-disk and thin-disk stars are not clearly distinguishable. Stars with no information of C or O abundances were not calculated. The observation parameters and element abundances of our sample stars are presented in Table \ref{tab:obs}.

\section{Stellar Models}

\subsection{Metal Mixtures}

The metal mixtures are used to construct opacity tables and to modify chemical compositions in stellar models. The high-temperature opacity tables are OPAL opacities constructed online. \footnote{http://opalopacity.llnl.gov/new.html} The low-temperature opacity tables are reconstructed according to observed C and O abundances \citep{Ferguson2005}. We call our metal mixtures as "CO extreme mix".

Based on the GS98 solar mixture \citep{GS98}, we construct new metal mixtures by adding enhancement factors to the solar log $N$ values of C, O, and $\alpha$-elements (i.e., Ne, Mg, Si, S, Ca, and Ti) with stable log $N_{\rm{Fe}}$ ($N_i$ represents the volume density of the element $i$) in the same way as \citet{Ge2015}. The enhancement factors are the the approximations from the observed [C/Fe], [O/Fe], and [$\alpha$/Fe]. For stars with [Fe/H] $<$ -0.8, we adopt the metal mixtures along the fitted line. For stars with -0.8 $<$ [Fe/H] $<$ -0.2, we use observed [O/Fe] and [C/Fe] for each star. We make an approximation for the $\alpha$-elements that high-$\alpha$ halo and thick-disk populations have [$\alpha$/Fe] = 0.3 dex, and the low-$\alpha$ halo stars have [$\alpha$/Fe] = 0.1. A few thick-disk stars in the metal-rich edge are with [$\alpha$/Fe] = 0.2, since the $\alpha$ abundances begin to decrease in the metal-rich edge. The metal mixtures we adopted are presented in Table \ref{tab:mix}.


We did not adopt the $\alpha$-enhancement directly from observations but made an approximation, because C and O take larger part in metal mixtures than the other elements. In Appendix A.2., we compare estimation parameters from models that have the same C and O abundances, but different $\alpha$-enhancements. The results indicate that the age differences would be similar or smaller than the estimation uncertainties, which means that the parameter differences estimated by different $\alpha$-elements can be ignored.


For scaled solar stars, the relationship between [Fe/H] and the ratio of the surface metal-element abundance to hydrogen abundance ($Z/X$) is $\log$ ($Z/X$) = $\log$ ($Z/X$)$_\odot$ + [Fe/H], where ($Z/X$)$_\odot$ is the ratio of the metal element to hydrogen for scaled solar mixture. For $\alpha$-enhanced stars, $Z$ values are always corrected by $f_{\alpha}$,
$Z$ = $Z_{0}$(0.694$f_{\alpha}$ + 0.306),
where $f_{\alpha}$ is the $\alpha$-enhancement factor and $Z_0$ is the overall metal abundance for a scaled solar metal mixture with the same [Fe/H]. These methods are not adequate for special metal mixtures, therefore we use the number fraction, mass fraction, and molecular weight of the elements to calculate $Z$ from [Fe/H] \citep{Ge2015}. Oxygen and carbon contribute most to $Z$, that is to say, with similar observed [Fe/H], metal mixtures with higher [O/Fe] or [C/Fe] give higher values of $Z$.

\subsection{Input Physics}

We compute a grid of evolutionary tracks using the Yale Rotation and Evolution Code (YREC, \citealt{Guenther1992}; \citealt{YangBi2007}; \citealt{Demarque2008}), in order to estimate the stellar parameters. The helium abundance obeys the Galactic helium enrichment, which is calibrated by standard solar models, $Y$ = 0.248 + 1.3324Z. The standard big bang nucleosynthesis value used in this equation is 0.248, following \citet{Spergel2007}. The mixing-length parameter $\alpha_{\ell}$ is fixed to 1.75. The models are calculated using the updated OPAL equation-of-state tables EOS2005 \citep{Rogers2002}. According to \citet{JW2011}, atomic diffusion is necessary in stellar models to estimate age values of the field halo stars, especially the diffusion of helium. Thus all models include gravitational settling of helium and heavy elements using the formulation of \citet{Thoul1994}.

\begin{figure}
\figurenum{1}
\plotone{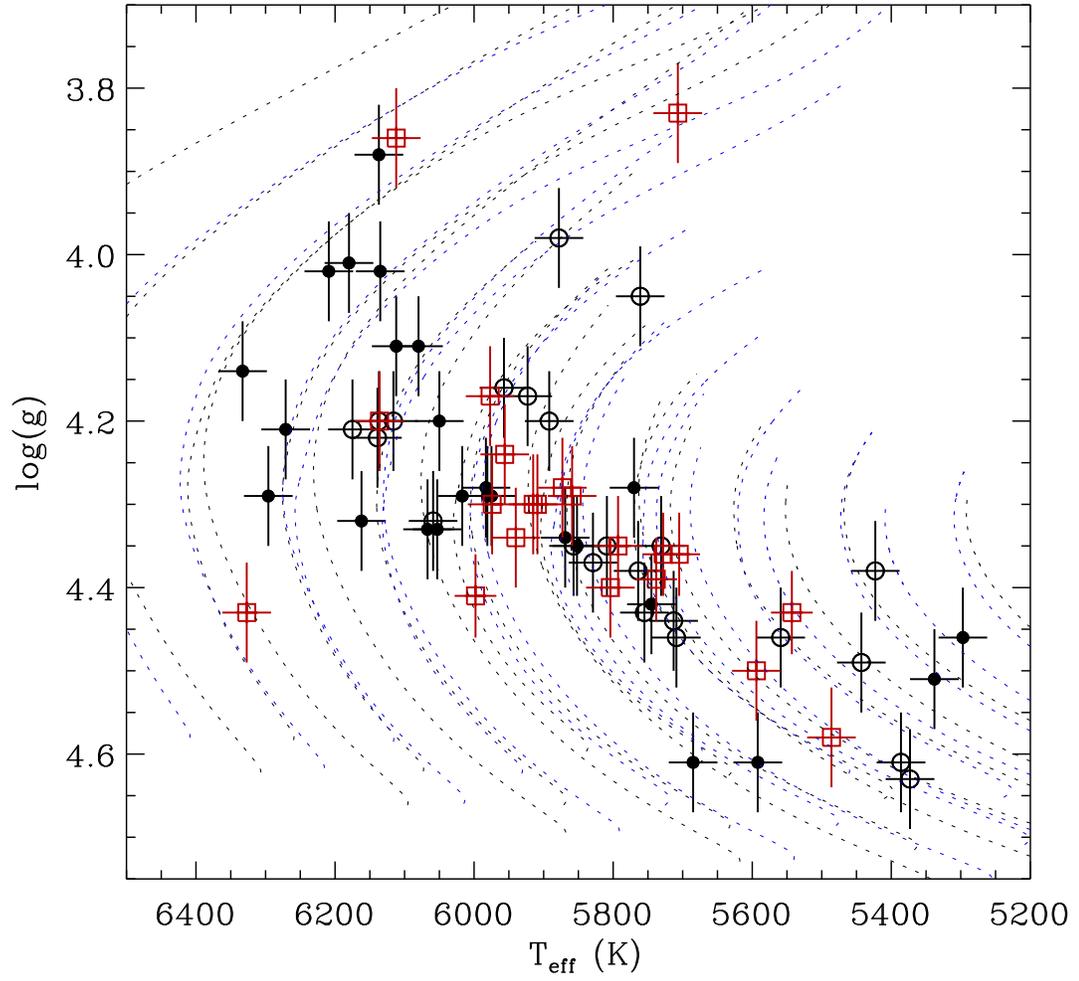}
\caption{The log $g$ - $T_{\rm{\rm{eff}}}$ diagram for the three populations. The open circles represent high-$\alpha$ halo stars, the filled circles are low-$\alpha$ halo stars. The red squares are thick-disk stars. The black dashed lines are evolution tracks with [C/Fe] = -0.1 dex, [O/Fe] = 0.3 dex and [$\alpha$/Fe] = 0.1 dex. The green lines are with [C/Fe] = 0.3 dex, [O/Fe] = 0.7 dex and [$\alpha$/Fe] = 0.3 dex. All the tracks are calculated from 0.60 M$_{\odot}$ to 0.90 M$_{\odot}$ with mass steps of 0.05 M$_{\odot}$. The [Fe/H] range is from -0.5 to -1.5 with a step of 0.5 dex. \label{fig:HR}}
\end{figure}

We calculate several evolution tracks that cover the observation constraints for each sample star. The mass range for each star is at 0.6 - 1.2 M$_\odot$. The mass step of the model grid is 0.01 M $_\odot$. The metallicity range is from -1.6 to -0.2 dex, the step is 0.1 dex for models with [Fe/H] $<$ -1.0, and 0.05 dex for those with [Fe/H] $>$ -1.0. The position of the stars in the log $g$-$T_{\rm{\rm{eff}}}$ diagram is shown in Figure \ref{fig:HR}.

\subsection{Parameter Estimation}

To estimate fundamental parameters for the sample stars, we select candidate models that fit observation constraints, i.e., log $g$, $T_{\rm{\rm{eff}}}$, and [Fe/H]. We follow the approach of \citet{Basu2010} and search for the most probable model in a Bayesian sense to determine the fundamental parameters (see also \citealt{Kallinger2010}).

With the Bayesian approach, we can identify the overall probability of the model $M_i$ with posterior probability $I$ fitting the observations property $D$ with respect to the whole selected models, according to the Bayes theorem,
\begin{equation}
p\left( {M_i \left| {D,I} \right.} \right) = \frac{{p\left( {M_i \left| I \right.} \right)p\left( {D\left| {M_i ,I} \right.} \right)}}{{P\left( {D\left| I \right.} \right)}}
  \label{0}
\end{equation}
where
\begin{equation}
p\left( {M_i \left| I \right.} \right) = \frac{1}{{N_m }}
  \label{1}
\end{equation}
is the uniform prior probability for a specific model with $N_m$ being the total number of selected models, and
\begin{equation}
p\left( {D\left| {M_i ,I} \right.} \right) = L\left( {T_{\rm{eff}} ,\log g,[{\rm{Fe}}/{\rm{H}}]} \right) = L_{T_{\rm{eff}} } L_{\log g} L_{[{\rm{Fe}}/{\rm{H}}]}
  \label{2}
\end{equation}
is the likelihood function. We use $T_{\rm{\rm{eff}}}$, log $g$, and [Fe/H] as observation constraints. In Equation \ref{0}, \begin{math}{P\left( {D\left| I \right.} \right)}\end{math} is a normalization factor for the specific model probability as follows:
\begin{equation}
P\left( {D\left| I \right.} \right) = \sum\limits_{j = 1}^{N_m } {p\left( {M_j \left| I \right.} \right) \cdot p\left( {D\left| {M_j ,I} \right.} \right)}
  \label{3}
\end{equation} The uniform priors (Equation \ref{1}) can be canceled, and we obtain the simplified Equation \ref{0} as follows:
\begin{equation}
p\left( {M_i \left| {D,I} \right.} \right) = \frac{{p\left( {D\left| {M_i ,I} \right.} \right)}}{{\sum\limits_{j = 1}^{N_m } {p\left( {D\left| {M_j ,I} \right.} \right)} }}.
  \label{4}
\end{equation} Then Equation \ref{4} is the probability distribution for the selected models with the most probable fundamental parameters. The uncertainties of the parameters are obtained by constructing the marginal distribution for each parameter.

The median value of each parameter is given with a probability $P$ = 0.5 in Equation \ref{4}. We adopt a 1$\sigma$ error for all the fundamental parameters, which means the low value of the parameter is with $P$ = 0.16, the high value is with $P$ = 0.84. Most of the stars have estimated age errors smaller than 1.0 Gyr. 10 out of 70 stars that are zero-age main-sequence (ZAMS) stars or close to ZAMS stars have age uncertainties larger than 2.0 Gyr. Table \ref{tab:estimation} lists the fundamental parameters estimated by the CO extreme mix models for all the sample stars.

\subsection{Comparison with $Y^2$ results}

To detect the influence of oxygen and carbon in stellar models and compare the CO extreme mix models with traditional $\alpha$-enhanced models, we chose $Y^2$ isochrones, which are also based on the YREC code and include diffusion, to estimate stellar parameters with observation constraints from \citet{Nissen14}. The $Y^2$ isochrones include helium diffusion \citep{Thoul1994} and convective core overshooting. The mixing-length parameter is 1.7431. The initial helium abundance is $Y$ = 0.23 + 2.0$Z$. The OPAL equation of state is adopted. The solar mixture used is from \citet{GN93}. The $\alpha$-enhanced pattern is similar to that of \citet{VandenBerg2000}, with O and $\alpha$-elements enhanced by the same factor. The opacity tables are OPAL high-temperature opacities, and the low-temperature opacities are adopted from \citet{AF94}. We construct isochrones with the exact observed [$\alpha$/Fe] and [Fe/H] range. We then use the Bayesian method to estimate the fundamental parameters. $Y^2$ estimated ages and masses for all the sample stars are presented in Table \ref{tab:estimation}.

\begin{figure}
\figurenum{2}
\plotone{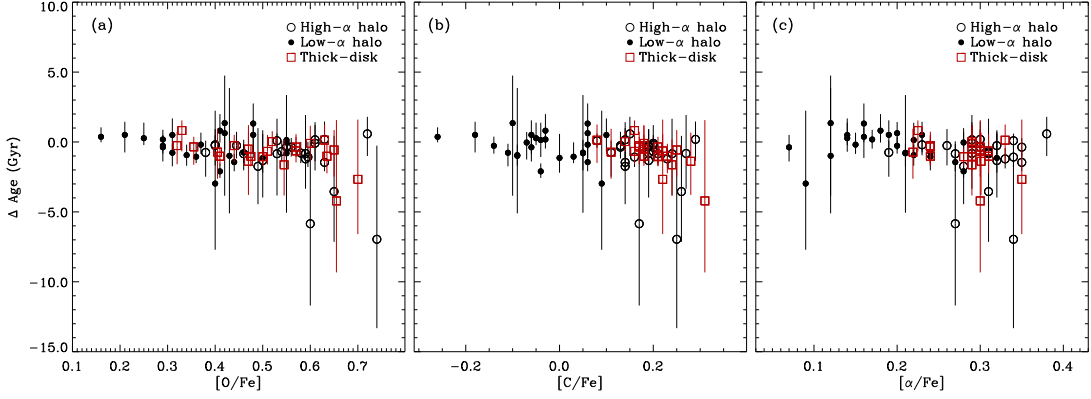}
\caption{Age difference between CO extreme mix results and $Y^2$ results against [O/Fe], [C/Fe] and [$\alpha$/Fe]. The y-coordinate is the CO extreme mix age minus the corresponding $Y^2$ age. The error of the $\Delta$ Age is contributed by the CO extreme mix age error and the $Y^2$ age error, using the error transfer formula \begin{math}\sigma  = \sqrt {\sigma _1^2  + \sigma _2^2 }\end{math}. \label{fig:AgeYY}}
\end{figure}

The systematic deviation between the two systems comes from input physics, for example, with or without overshooting, the different treatment of the diffusion, the initial helium abundance $Y$, mixing-length parameter, and metal mixtures used in the models. The age differences between CO extreme mix results and $Y^2$ results for all the sample stars are shown in Figure \ref{fig:AgeYY}. From this figure, we note that the age differences increase slightly when [O/Fe], [C/Fe] and [$\alpha$/Fe] are increased. Thus, the age differences between two systems might come from different metal mixtures. The $\alpha$-enhanced models treat oxygen to be enhanced by the same factor as $\alpha$-elements, because an observed O abundance is not always available. Moreover, the enhancement of C is not considered. Hence, traditional $\alpha$-enhanced models might be inadequate for the stars with highly enhanced or decreased O or C abundances. For stars with extremely high O and C abundances, traditional $\alpha$-enhanced mixtures would give a lower estimate for $Z$ and would determine a higher stellar mass, which means that they might overestimate the stellar ages.

\section{Ages of Dwarfs in the Three Populations}

\subsection{Age distribution}

\begin{figure}
\figurenum{3}
\plotone{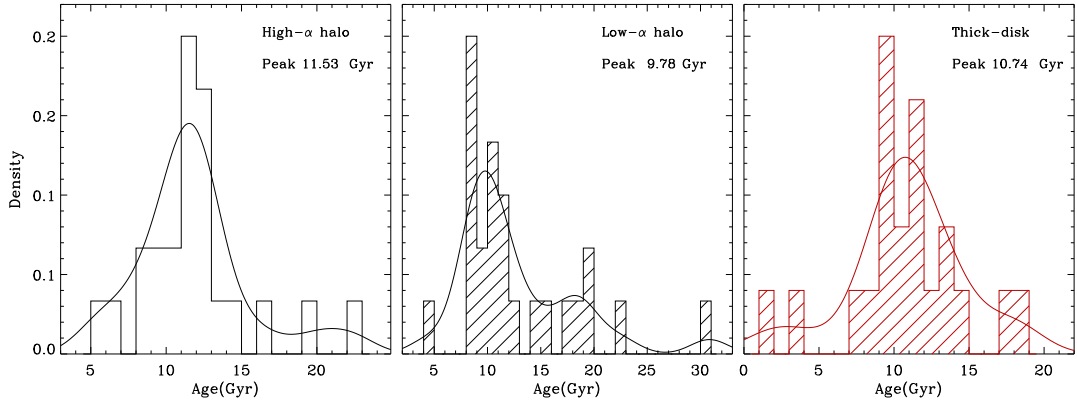}
\caption{Age distribution for the three populations. The y-coordinate represents the kernel density for stellar age. The solid lines delineate a nonparametric kernel density estimator (Gaussian smoothing) to the age distribution profiles. The histogram profiles for stellar ages have been normalized to a peak value of 0.2.} \label{fig:Agedist}
\end{figure}

The age distribution profiles of the three populations are shown in Figure \ref{fig:Agedist}. The age distribution profiles indicate that most of the halo stars are older than 8 Gyr, which agrees with the knowledge that the Galaxy halo was assembled very early. The high-$\alpha$ halo stars are condensed in the range of 11 - 13 Gyr, the low-$\alpha$ halo stars are focused at around 8 - 11 Gyr, and the thick-disk stars are located at about 9 - 14 Gyr. As a result, the mean age of the high-$\alpha$ halo population is older by about 2 $\sim$ 3 Gyr than that of the low-$\alpha$ halo stars; the thick-disk stars have medium ages. Our results present the same relative age of the two halo populations as reported by \citet{Schuster2012}, and we confirm their prediction about the ages of the thick-disk stars. We perform a nonparametric kernel density estimator of the age distribution to define the peak value of the age distribution. The difference in the peak of the age distribution between high-$\alpha$ halo and low-$\alpha$ halo stars is about 2 Gyr.

The high- and low-$\alpha$ halo stars have different age distribution profiles. The age range of the high-$\alpha$ halo is relatively narrow. The low-$\alpha$ halo stars are younger, and many sample stars divert from the center age. The age distribution characters of the two halo populations fit their observed chemistry behaviors. The C and O abundances of the high-$\alpha$ halo stars have smaller star-to-star scatter than the abundances of low-$\alpha$ halo stars. Thus high-$\alpha$ halo stars have age values more concentrated than those of low-$\alpha$ halo stars. These features also reflect the formation and evolution process of the two halo populations. The high-$\alpha$ halo stars were possibly formed in an environment where SNe II dominated the chemical composition of the interstellar medium; their formation timescales are relatively short. The low-$\alpha$ halo stars were probably born in accreted gas mixed with different outside galaxies; their ages are dispersed in a wide range. The age distribution profiles of the two halo populations seem to fit the in situ accretion models of \citet{Zolotov2009,Zolotov2010}.

The disk stars have the flattest age distribution profile and show little scatter. This might indicate that these stars were formed in a very gentle process. From chemical compositions, we could not distinguish the high-$\alpha$ halo and thick-disk stars. It seems that they were born in a similar environment, where massive stars that explode as SNe II dominate the interstellar medium. However, from the width of the age distribution profiles, we could say that the thick-disk stars may have a relatively longer star formation process. Based on this, we could draw a picture that the ancient in-situ halo was formed during the collapse, which was a relatively severe process. After this, the thick-disk stars started to be constructed in a relatively quiescent and long-lasting process.

\subsection{Age-chemistry ([O/Fe], [C/Fe], and [Fe/H])}

\begin{figure*}
\figurenum{4}
\centering
\includegraphics[width=18.0cm]{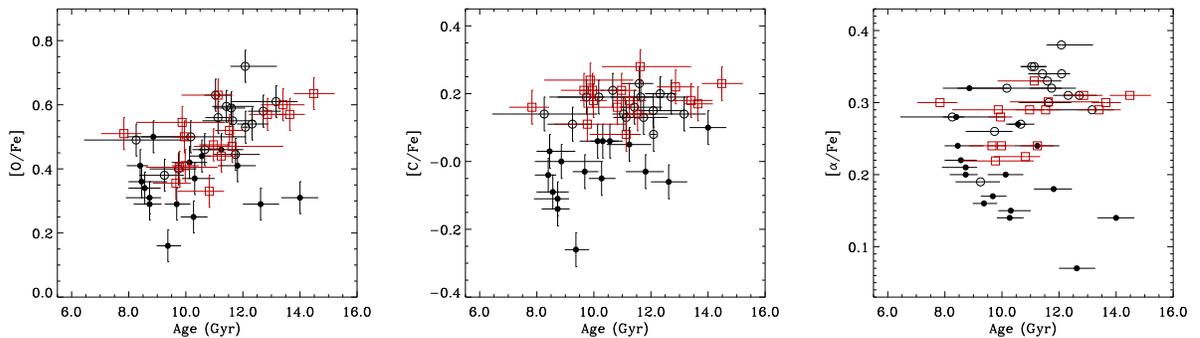}
   \caption{Age vs. [O/Fe], [C/Fe], and [$\alpha$/Fe]. The stars are in the age range of 7.0-15.0 Gyr. The samples with age errors larger than 2.0 Gyr have been deleted. The black solid dots are low-$\alpha$ halo stars, black circles present high-$\alpha$ halo stars, and red squares are thick-disk stars. }
\label{fig:AgeCOA}
\end{figure*}

We delete samples with large estimation age errors ($\sigma$ $>$ 2.0 Gyr), since most of them depart from the behaviors of the majority samples. According to the Universe age and the ages of the most ancient halo stars (e.g., \citealt{Bennett2013}; \citealt{VandenBerg2014}), we select stars with age $<$ 15.0 Gyr and exclude the very young samples from the chemistry analysis. Figure \ref{fig:AgeCOA} shows the ages against the [O/Fe], [C/Fe], and [$\alpha$/Fe]. There seems to be a trend that stars with higher [O/Fe] or [C/Fe] are older; most of the high-$\alpha$ halo stars and thick-disk stars are older than low-$\alpha$ halo stars. Stellar ages and O, C abundances fit a linear relation, which means that O and C abundances are good indicators for stellar ages. The age-[O/Fe] profile has the smallest scatter; it seems that stellar ages are the most sensitive to the O abundance.

\begin{figure}
\figurenum{5}
\plotone{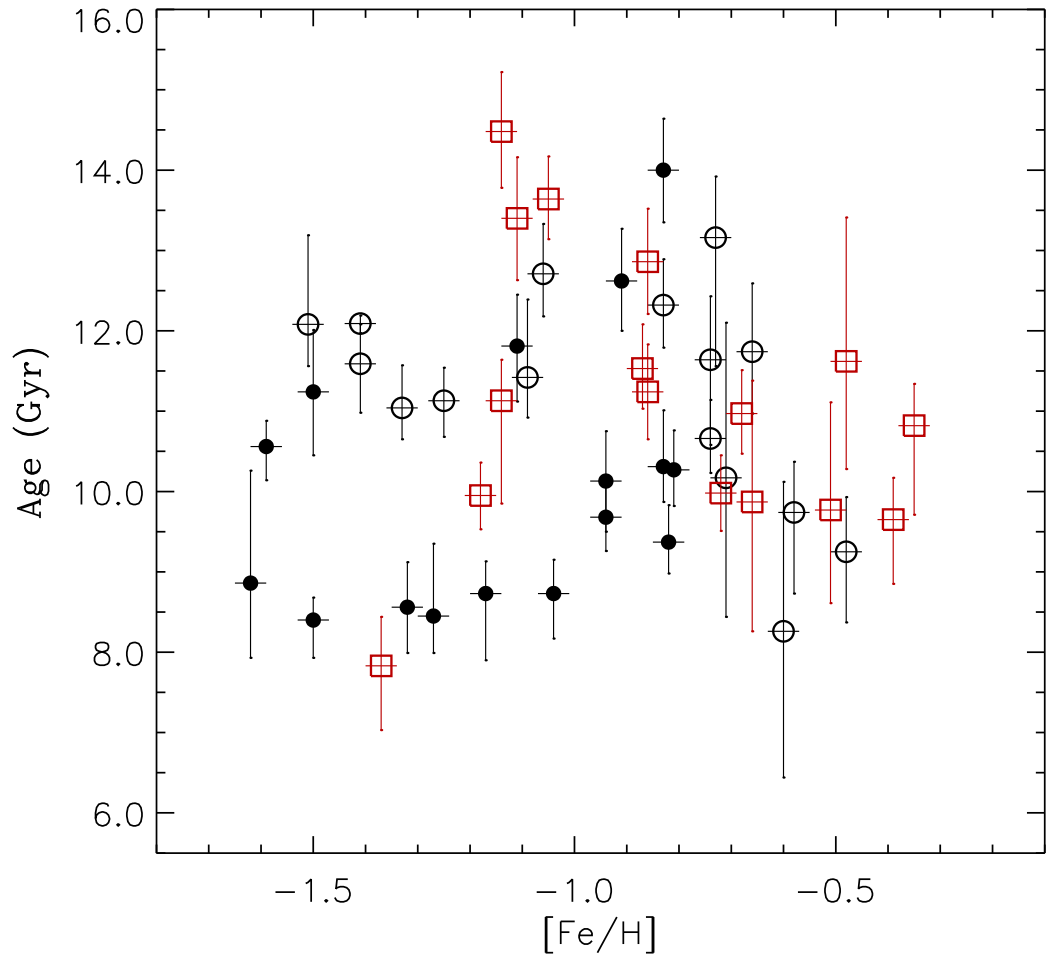}
\caption{Age-metallicity relation. The stars have ages from 7.0 Gyr to 15.0 Gyr. The samples with age errors larger than 2.0 Gyr have been deleted. The symbols are the same as in Figure \ref{fig:AgeCOA}. \label{fig:Agefeh}}
\end{figure}

We also describe the age - metallicity relation for the three populations in Figure \ref{fig:Agefeh}. The halo stars are divided into two populations, i.e., the young low-$\alpha$ halo and the old high-$\alpha$ halo. According to the mean age and median age of the two populations, the age difference would be 1.0 Gyr at the metal-poor edge. The ages of high-$\alpha$ halo and thick-disk stars decline with increasing metallicity and have larger scatter in the metal-poor edge. This behavior is probably due to the observation properties of O and C abundances of the stars: that O and C abundances are dispersed more widely at the metal-poor edge.

\subsection{Stars with peculiar ages}

HD 106516 has obvious thick-disk population chemical and dynamical characters ([$\alpha$/Fe] = 0.29 dex, [C/Fe] = 0.18 dex, [O/Fe] = 0.47 dex, [Fe/H] = - 0.69; V$_{total}$ = 100 kms$^{-1}$). However, our results and many works have confirmed that it is a very young star with rotation and strong activity (e.g., \citealt{Barnes2007}; \citealt{Schroder2013}; \citealt{Sitnova2015}). Moreover, \citet{Carney2001} regarded it as a single-lined spectroscopic binary, which means that it has a very dim or far away companion object. We could guess that most materials of the companion object have been accreted by HD 106516. This companion star may present a high-$\alpha$ chemical composition and influence the atmosphere of HD 106516. Furthermore, because of the accretion, the dynamics of HD 106516 have been heated, which cause it to resemble a thick-disk star.

HD 65907 is selected as a thick-disk star based on its chemical composition, [$\alpha$/Fe] = 0.24, [O/Fe] = 0.32, [C/Fe] = 0.17. However, we estimate it to be a young star with $t$ = 3.7 $\sim$ 4.0 Gyr. Previous works estimated its age at around 4.5 Gyr - 9.6 Gyr (e.g., \citealt{RP1998}; \citealt{Nordstrom2004}; \citealt{VF2005}). In addition, when checking the dynamic parameters, U = 23 km$^{-1}$, V = -12km$^{-1}$, W = 41km$^{-1}$, \citet{Adibekyan2012} regarded it as a thin-disk star. In fact, HD 65907 lies in the metal-rich region ([Fe/H] = -0.33), where thick-disk and thin-disk stars are contaminated. Thus, we assume that HD 65907 might be a transit-disk star.

We also obtained twelve stars that are much older than the age of the Universe ($t$ $>$ 15 Gyr). Among these stars, CD-610282, G05-19, G53-41, HD 193901 were determined with ages in the range of 12 - 15 Gyr by \citet{Schuster2012} with $Y^2$ isochrones, but our results give age values in the range of 15.5 - 22.5 Gyr obtained with CO extreme mix models, and 15.0 - 22.0 Gyr by $Y^2$ isochrones. Using observation parameters from \citet{Schuster2012} and $Y^2$ isochrones, we obtain the age values of 11.0 - 15.0 Gyr for these stars. Thus the age differences between our results and those by \citet{Schuster2012} mainly come from different observation properties, i.e., T$_{\rm{eff}}$, log $g$, and [Fe/H].

It seems that there were more particular old stars in the low-$\alpha$ halo population than in the other two populations (two thick-disk stars, three high-$\alpha$ halo stars, seven low-$\alpha$ halo stars). Maybe because low-$\alpha$ halo stars are formed in the materials accreted from ancient dwarf galaxies or globular clusters, it is easier to find very old samples in the low-$\alpha$ halo population. The origin of the extremely old stars from our sample is still not clear, but these stars need further studies. With more observation constraints, the ages of these stars might be measured more accurately and may be candidates for clocks of the universe.

\section{Conclusions}

We obtained more accurate and self-consistent stellar ages for the 70 metal-poor dwarf stars, including C and O abundances in stellar models. The age distribution profiles show that the high-$\alpha$ halo population is somewhat older than the low-$\alpha$ halo population by around 2.0 Gyr. The thick-disk population has an age range in between the high- and low-$\alpha$ halo populations. The two halo populations match the in situ accretion halo models. The thick-disk stars share a similar chemistry with the high-$\alpha$ halo population, but have a flatter age distribution profile, which means that these thick-disk stars may be born in a relatively stable environment with a long formation time scale.

The relationships of ages and [O/Fe], [C/Fe] reflect that O and C abundances are important indicators of the stellar ages. The relation between ages and [O/Fe] also show the smallest scatter, which means that the stellar age is the most sensitive to the O abundance. The age-metallicity distribution describes the age difference between high-$\alpha$ halo and low-$\alpha$ halo stars at the metal-poor edge. Furthermore, the trend of the ages against the [Fe/H] coincides with the observed O and C abundances.

We found two very young thick-disk stars as well as some stars older than the universe. The peculiar ages of these samples led us to investigate the populations of these stars with more caution.

\acknowledgments

This work is supported by grants 11273007 and 10933002 from the National Natural Science Foundation of China, the Joint Research Fund in Astronomy (U1631236) under cooperative agreement between the National Natural Science Foundation of China (NSFC) and Chinese Academy of Sciences (CAS), and the Fundamental Research Funds for the Central Universities. This work is also supported by the Youth Scholars Program of Beijing Normal University.

\clearpage

\clearpage

\begin{deluxetable}{ccccccccccccc}
\tablecaption{Atmospheric Parameters and Element abundances for the stars. \label{tab:obs}}
\tablehead{
ID               &  $T_{\rm{eff}}$   &   $\log$ g  & [Fe/H]  & [$\alpha$/Fe]   &  [C/H]            &[O/H]$_{7774}$  &  [O/H]$_{6300}$    &  [C/Fe] &    [O/Fe]        & Pop. \\
& K & dex  &   &   & nLTE  & nLTE  & nLTE  &   &
}
 \renewcommand{\arraystretch}{1.5}
\startdata
CD-436810	&	6059	&	4.32 	&	-0.44 	&	0.23 	&	-0.24 	 &	 -0.08 	&	 0.00 	&	0.20 	&	0.44 	&	A	\\
G05-36	&	6139	&	4.22 	&	-1.25 	&	0.35 	&	-1.12 	&	 -0.69 	&	...	 &	1.25 	&	1.25 	&	A	\\
G05-40	&	5892	&	4.20 	&	-0.83 	&	0.31 	&	-0.63 	&	 -0.29 	&	...	 &	0.83 	&	0.83 	&	A	\\
G15-23	&	5373	&	4.63 	&	-1.12 	&	0.34 	&	-0.87 	&	 -0.38 	&	...	 &	1.12 	&	1.12 	&	A	\\
G18-28	&	5443	&	4.49 	&	-0.85 	&	0.31 	&	-0.59 	&	 -0.20 	&	...	 &	0.85 	&	0.85 	&	A	\\
G18-39	&	6175	&	4.21 	&	-1.41 	&	0.34 	&	-1.33 	&	 -0.88 	&	...	 &	1.41 	&	1.41 	&	A	\\
G24-13	&	5764	&	4.38 	&	-0.73 	&	0.29 	&	-0.59 	&	 -0.12 	&	...	 &	0.73 	&	0.73 	&	A	\\
G31-55	&	5731	&	4.35 	&	-1.12 	&	0.29 	&	-0.92 	&	 -0.51 	&	...	 &	1.12 	&	1.12 	&	A	\\
G85-13	&	5709	&	4.46 	&	-0.60 	&	0.28 	&	-0.46 	&	 -0.11 	&	...	 &	0.60 	&	0.60 	&	A	\\
G99-21	&	5559	&	4.46 	&	-0.68 	&	0.29 	&	-0.41 	&	 -0.09 	&	...	 &	0.68 	&	0.68 	&	A	\\
G159-50	&	5713	&	4.44 	&	-0.94 	&	0.31 	&	-0.70 	&	 -0.41 	&	...	 &	0.94 	&	0.94 	&	A	\\
G180-24	&	6137	&	4.20 	&	-1.41 	&	0.33 	&	-1.18 	&	 -0.82 	&	...	 &	1.41 	&	1.41 	&	A	\\
G188-22	&	6116	&	4.20 	&	-1.33 	&	0.35 	&	-1.19 	&	 -0.70 	&	...	 &	1.33 	&	1.33 	&	A	\\
HD 51754	&	5857	&	4.35 	&	-0.58 	&	0.26 	&	-0.39 	&	 -0.17 	&	 -0.19 	&	0.58 	&	0.58 	&	A	\\
HD 111980	&	5878	&	3.98 	&	-1.09 	&	0.34 	&	-0.93 	 &	 -0.46 	&	 -0.53 	&	1.09 	&	1.09 	&	A	\\
HD 113679	&	5761	&	4.05 	&	-0.66 	&	0.32 	&	-0.53 	 &	 -0.19 	&	 -0.24 	&	0.66 	&	0.66 	&	A	\\
HD 121004	&	5755	&	4.43 	&	-0.71 	&	0.32 	&	-0.52 	 &	 -0.22 	&	 -0.20 	&	0.71 	&	0.71 	&	A	\\
HD 132475	&	5750	&	3.77 	&	-1.51 	&	0.38 	&	-1.36 	 &	 -0.79 	&	...	 &	1.51 	&	1.51 	&	A	\\
HD 148816	&	5923	&	4.17 	&	-0.74 	&	0.27 	&	-0.53 	 &	 -0.28 	&	...	 &	0.74 	&	0.74 	&	A	\\
HD 159482	&	5829	&	4.37 	&	-0.74 	&	0.30 	&	-0.55 	 &	 -0.19 	&	...	 &	0.74 	&	0.74 	&	A	\\
HD 160693	&	5809	&	4.35 	&	-0.48 	&	0.19 	&	-0.37 	 &	 -0.10 	&	...	 &	0.48 	&	0.48 	&	A	\\
HD 179626	&	5957	&	4.16 	&	-1.06 	&	0.31 	&	-0.87 	 &	 -0.48 	&	...	 &	1.06 	&	1.06 	&	A	\\
HD 222766	&	5423	&	4.38 	&	-0.70 	&	0.30 	&	-0.41 	 &	 -0.07 	&	...	 &	0.70 	&	0.70 	&	A	\\
HD 230409	&	5386	&	4.61 	&	-0.87 	&	0.27 	&	-0.70 	 &	 -0.27 	&	...	 &	0.87 	&	0.87 	&	A	\\
\noalign{\smallskip}\hline\noalign{\smallskip}																																	 CD-453283	&	5685	&	4.61 	&	-0.93 	&	0.12 	&	-1.02 	 &	 -0.50 	&	...	 &	-0.09 	&	0.43 	&	B	\\
CD-514628	&	6296	&	4.29 	&	-1.32 	&	0.22 	&	-1.41 	 &	 -0.98 	&	...	 &	-0.09 	&	0.34 	&	B	\\
CD-571633	&	5981	&	4.29 	&	-0.91 	&	0.07 	&	-0.97 	 &	 -0.62 	&	...	 &	-0.06 	&	0.29 	&	B	\\
CD-610282	&	5869	&	4.34 	&	-1.25 	&	0.22 	&	-1.29 	 &	 -0.70 	&	...	 &	-0.04 	&	0.55 	&	B	\\
G05-19	&	5770	&	4.28 	&	-1.19 	&	0.19 	&	-1.25 	&	 -0.71 	&	...	 &	-0.06 	&	0.48 	&	B	\\
G20-15	&	6162	&	4.32 	&	-1.50 	&	0.24 	&	-1.45 	&	 -1.04 	&	...	 &	0.05 	&	0.46 	&	B	\\
G46-31	&	6017	&	4.29 	&	-0.83 	&	0.15 	&	-0.77 	&	 -0.51 	&	 -0.41	&	0.06 	&	0.37 	&	B	\\
G53-41	&	5975	&	4.29 	&	-1.21 	&	0.23 	&	-1.39 	&	 -1.00 	&	...	 &	-0.18 	&	0.21 	&	B	\\
G56-36	&	6067	&	4.33 	&	-0.94 	&	0.20 	&	-0.88 	&	 -0.52 	&	...	 &	0.06 	&	0.42 	&	B	\\
G66-22	&	5297	&	4.46 	&	-0.88 	&	0.12 	&	-0.98 	&	 -0.46 	&	...	 &	-0.10 	&	0.42 	&	B	\\
G75-31	&	6135	&	4.02 	&	-1.04 	&	0.20 	&	-1.18 	&	 -0.75 	&	...	 &	-0.14 	&	0.29 	&	B	\\
G82-05	&	5338	&	4.51 	&	-0.78 	&	0.09 	&	-0.69 	&	 -0.38 	&	...	 &	0.09 	&	0.40 	&	B	\\
G112-43	&	6209	&	4.02 	&	-1.27 	&	0.24 	&	-1.24 	&	 -0.91 	&	...	 &	0.03 	&	0.36 	&	B	\\
G119-64	&	6333	&	4.14 	&	-1.50 	&	0.28 	&	-1.54 	&	 -1.09 	&	...	 &	-0.04 	&	0.41 	&	B	\\
G150-40	&	6080	&	4.11 	&	-0.82 	&	0.16 	&	-1.08 	&	 -0.66 	&	...	 &	-0.26 	&	0.16 	&	B	\\
G170-56	&	6112	&	4.11 	&	-0.94 	&	0.17 	&	-0.97 	&	 -0.65 	&	...	 &	-0.03 	&	0.29 	&	B	\\
HD 3567	&	6180	&	4.01 	&	-1.17 	&	0.21 	&	-1.28 	&	 -0.86 	&	...	 &	-0.11 	&	0.31 	&	B	\\
HD 59392	&	6137	&	3.88 	&	-1.62 	&	0.32 	&	-1.62 	&	 -1.12 	&	...	 &	0.00 	&	0.50 	&	B	\\
HD 103723	&	6050	&	4.20 	&	-0.81 	&	0.14 	&	-0.86 	 &	 -0.56 	&	...	 &	-0.05 	&	0.25 	&	B	\\
HD 105004	&	5852	&	4.35 	&	-0.83 	&	0.14 	&	-0.73 	 &	 -0.52 	&	...	 &	0.10 	&	0.31 	&	B	\\
HD 163810	&	5592	&	4.61 	&	-1.22 	&	0.21 	&	-1.17 	 &	 -0.67 	&	...	 &	0.05 	&	0.55 	&	B	\\
HD 193901	&	5745	&	4.42 	&	-1.11 	&	0.16 	&	-1.05 	 &	 -0.63 	&	...	 &	0.06 	&	0.48 	&	B	\\
HD 194598	&	6053	&	4.33 	&	-1.11 	&	0.18 	&	-1.14 	 &	 -0.70 	&	...	 &	-0.03 	&	0.41 	&	B	\\
HD 219617	&	5983	&	4.28 	&	-1.46 	&	0.28 	&	-1.53 	 &	 -0.91 	&	...	 &	-0.07 	&	0.55 	&	B	\\
HD 284248	&	6271	&	4.21 	&	-1.59 	&	0.27 	&	-1.53 	 &	 -1.15 	&	...	 &	0.06 	&	0.44 	&	B	\\
\noalign{\smallskip}\hline\noalign{\smallskip}
BD-213420	&	5909	&	4.30 	&	-1.14 	&	0.31 	&	-0.91 	 &	 -0.49 	&	 -0.52 	&	0.23 	&	0.64 	&	C	\\
CD-333337	&	6112	&	3.86 	&	-1.37 	&	0.30 	&	-1.21 	 &	 -0.86 	&	...	 &	0.16 	&	0.51 	&	C	\\
\textbf{HD 4308} 	&	5705	&	4.36 	&	-0.35 	&	0.23 	&	 -0.19 	 &	 -0.02 	& ...		&	0.16 	&	0.35 	&	C	\\
HD 17820	&	5873	&	4.28 	&	-0.68 	&	0.29 	&	-0.47 	&	 -0.20 	&	 -0.21 	&	0.21 	&	0.68 	&	C	\\
HD 22879	&	5859	&	4.29 	&	-0.86 	&	0.31 	&	-0.64 	&	 -0.29 	&	...	 &	0.22 	&	0.57 	&	C	\\
HD 25704	&	5974	&	4.30 	&	-0.86 	&	0.24 	&	-0.68 	&	 -0.44 	&	 -0.40 	&	0.18 	&	0.44 	&	C	\\
\textbf{HD 65907} 	&	5998	&	4.41 	&	-0.33 	&	0.24 	&	 -0.16 	 &	 -0.05 	&	0.03 	&	0.17 	&	0.33 	&	C \\
HD 76932	&	5977	&	4.17 	&	-0.87 	&	0.29 	&	-0.73 	&	 -0.40 	&	 -0.30 	&	0.14 	&	0.52 	&	C	\\
\textbf{HD 77110} 	&	5738	&	4.39 	&	-0.51 	&	0.22 	&	 -0.40 	 &	 -0.14 	&	-0.07 	&	0.11 	&	0.51 	&	C \\
HD 97320	&	6136	&	4.20 	&	-1.18 	&	0.28 	&	-0.97 	&	 -0.68 	&	...	 &	0.21 	&	0.50 	&	C	\\
HD 106516	&	6327	&	4.43 	&	-0.69 	&	0.29 	&	-0.51 	 &	 -0.22 	&	...	 &	0.18 	&	0.69 	&	C	\\
\textbf{HD 111232} 	&	5543	&	4.43 	&	-0.48 	&	0.30 	&	 -0.20 	 &	 -0.01 	&...		&	0.28 	&	0.48 	&	C	\\
HD 114762	&	5956	&	4.24 	&	-0.72 	&	0.24 	&	-0.54 	 &	 -0.31 	&	...	 &	0.18 	&	0.72 	&	C	\\
HD 120559	&	5486	&	4.58 	&	-0.91 	&	0.30 	&	-0.60 	 &	 -0.26 	&	 -0.25 	&	0.31 	&	0.66 	&	C	\\
HD 126681	&	5594	&	4.50 	&	-1.20 	&	0.35 	&	-0.98 	 &	 -0.50 	&	...	 &	0.22 	&	0.70 	&	C	\\
\textbf{HD136352} 	&	5728	&	4.36 	&	-0.39 	&	0.24 	&	 -0.18 	 &	 -0.07 	&	0.00 	&	0.21 	&	0.39 	&	C \\
HD 175179	&	5804	&	4.40 	&	-0.66 	&	0.29 	&	-0.42 	 &	 -0.09 	&	 -0.14 	&	0.24 	&	0.66 	&	C	\\
HD 189558	&	5707	&	3.83 	&	-1.14 	&	0.33 	&	-1.06 	 &	 -0.51 	&	...	 &	0.08 	&	0.63 	&	C	\\
HD 199289	&	5915	&	4.30 	&	-1.05 	&	0.30 	&	-0.88 	 &	 -0.49 	&	 -0.47 	&	0.17 	&	0.57 	&	C	\\
HD 205650	&	5793	&	4.35 	&	-1.19 	&	0.30 	&	-0.94 	 &	 -0.54 	&	...	 &	0.25 	&	0.65 	&	C	\\
HD 241253	&	5940	&	4.34 	&	-1.11 	&	0.29 	&	-0.93 	 &	 -0.51 	&	...	 &	0.18 	&	0.60 	&	C	\\
\enddata
\tablenotetext{}{Population classification: A, high-$\alpha$ halo; B, low-$\alpha$ halo; C, thick-disk population.}
\tablenotetext{}{The observation errors for $T_{\rm{eff}}$, log $g$, and [Fe/H] are 35 K, 0.06 dex, and 0.03 dex. The [$\alpha$/Fe] values are from \citet{NS10}. }
\tablenotetext{}{Stars in bold font have observation errors of 30 K, 0.05 dex, and 0.03 dex for T$_{\rm{eff}}$, log $g$, and [Fe/H]. The [$\alpha$/Fe] values are calculated by making an average of [Mg/Fe], [Si/Fe], and [Ti/Fe] from \citet{Adibekyan2012}.}
\end{deluxetable}

\begin{deluxetable}{cccc }
\tablecaption{The Metal Mixtures in Stellar Models \label{tab:mix}}
\tablehead{
 Metal Mix &  $[{\rm{C}}/{\rm{Fe}}]$  & $[{\rm{O}}/{\rm{Fe}}]$ & $[\alpha /{\rm{Fe}}]$
 }
\startdata
high-$\alpha$   & 0.3   & 0.7  & 0.3    \\
                        & 0.3   & 0.6  & 0.3   \\
                        & 0.2   & 0.6  & 0.3    \\
                        & 0.1   & 0.6  & 0.3  \\
                           & 0.3   & 0.5  & 0.3    \\
                           & 0.2   & 0.5  & 0.3    \\
                           & 0.1   & 0.5  & 0.3    \\
                        & 0.2   & 0.4  & 0.2    \\
                        & 0.1   & 0.4  & 0.2    \\
                     \noalign{\smallskip}\hline\noalign{\smallskip}
       low-$\alpha$    & 0.1   & 0.5  & 0.1    \\
                       & 0.0   & 0.4  & 0.1    \\
                       &-0.1   & 0.3  & 0.1    \\
\enddata
\end{deluxetable}

\clearpage

\begin{deluxetable}{ccccccccccccc}
\tablecaption{Fundamental parameters of the stars. \label{tab:estimation}}
\tablehead{
ID & T$_{\rm{eff}}$& log $g$ & Age$_{CO - mix}$ & Mass$_{CO - mix}$ & Age$_{Y^2}$  & Mass$_{Y^2}$  & Pop.  \\
 & K & dex & Gyr & $M_{\odot}$  & Gyr & $M_{\odot}$
}
\renewcommand{\arraystretch}{1.5}
\startdata
CD-436810	&	$	6059	~_{-~	23	}^{+~	25	}	$	&	$	 4.31 	 ~_{-~	0.03	}^{+~	0.04 	}	$	&	$	5.30 	~_{-~	 0.75	 }^{+~	 0.47 	}	$	&	$	1.03 	~_{-~	0.02	}^{+~	 0.01 	}	 $	&	$	5.50 	~_{-~	0.00	}^{+~	0.50 	}	$	 &	$	 1.00 	 ~_{-~	0.01	 }^{+~	0.00 	}	$	&	A	\\
G05-36	&	$	6142	~_{-~	21	}^{+~	13	}	$	&	$	4.24 	 ~_{-~	0.04	}^{+~	0.02 	}	$	&	$	11.13 	~_{-~	0.45	 }^{+~	0.41 	 }	$	&	$	0.80 	~_{-~	0.00	}^{+~	0.01 	 }	 $	&	$	11.50 	~_{-~	0.50	}^{+~	0.50 	}	$	&	$	 0.82 	 ~_{-~	0.00	}^{+~	 0.01 	}	$	&	A	\\
G05-40	&	$	5893	~_{-~	23	}^{+~	22	}	$	&	$	4.21 	 ~_{-~	0.04	}^{+~	0.04 	}	$	&	$	12.32 	~_{-~	0.53	 }^{+~	0.57 	 }	$	&	$	0.84 	~_{-~	0.01	}^{+~	0.01 	 }	 $	&	$	13.00 	~_{-~	0.50	}^{+~	0.50 	}	$	&	$	 0.84 	 ~_{-~	0.02	}^{+~	 0.01 	}	$	&	A	\\
G15-23	&	$	5374	~_{-~	23	}^{+~	21	}	$	&	$	4.64 	 ~_{-~	0.04	}^{+~	0.03 	}	$	&	$	8.04 	~_{-~	4.17	 }^{+~	5.05 	 }	$	&	$	0.68 	~_{-~	0.22	}^{+~	0.02 	 }	 $	&	$	15.00 	~_{-~	4.80	}^{+~	4.40 	}	$	&	$	 0.66 	 ~_{-~	0.01	}^{+~	 0.01 	}	$	&	A	\\
G18-28	&	$	5444	~_{-~	22	}^{+~	20	}	$	&	$	4.50 	 ~_{-~	0.04	}^{+~	0.04 	}	$	&	$	16.46 	~_{-~	2.58	 }^{+~	3.45 	 }	$	&	$	0.71 	~_{-~	0.19	}^{+~	0.01 	 }	 $	&	$	20.01 	~_{-~	2.50	}^{+~	2.00 	}	$	&	$	 0.69 	 ~_{-~	0.01	}^{+~	 0.01 	}	$	&	A	\\
G18-39	&	$	6162	~_{-~	16	}^{+~	16	}	$	&	$	4.26 	 ~_{-~	0.01	}^{+~	0.01 	}	$	&	$	12.09 	~_{-~	0.12	 }^{+~	0.10 	 }	$	&	$	0.78 	~_{-~	0.00	}^{+~	0.00 	 }	 $	&	$	12.00 	~_{-~	0.50	}^{+~	0.50 	}	$	&	$	 0.80 	 ~_{-~	0.01	}^{+~	 0.01 	}	$	&	A	\\
G24-13	&	$	5766	~_{-~	21	}^{+~	8	}	$	&	$	4.35 	 ~_{-~	0.02	}^{+~	0.04 	}	$	&	$	13.16 	~_{-~	1.60	 }^{+~	0.76 	 }	$	&	$	0.81 	~_{-~	0.00	}^{+~	0.01 	 }	 $	&	$	13.00 	~_{-~	1.50	}^{+~	1.00 	}	$	&	$	 0.80 	 ~_{-~	0.01	}^{+~	 0.01 	}	$	&	A	\\
G31-55	&	$	5734	~_{-~	20	}^{+~	14	}	$	&	$	4.35 	 ~_{-~	0.03	}^{+~	0.03 	}	$	&	$	19.91 	~_{-~	0.82	 }^{+~	0.83 	 }	$	&	$	0.70 	~_{-~	0.00	}^{+~	0.01 	 }	 $	&	$	20.00 	~_{-~	1.00	}^{+~	1.00 	}	$	&	$	 0.70 	 ~_{-~	0.01	}^{+~	 0.01 	}	$	&	A	\\
G85-13	&	$	5710	~_{-~	22	}^{+~	21	}	$	&	$	4.47 	 ~_{-~	0.04	}^{+~	0.03 	}	$	&	$	8.26 	~_{-~	1.82	 }^{+~	1.86 	 }	$	&	$	0.85 	~_{-~	0.01	}^{+~	0.01 	 }	 $	&	$	10.00 	~_{-~	2.00	}^{+~	2.00 	}	$	&	$	 0.83 	 ~_{-~	0.01	}^{+~	 0.01 	}	$	&	A	\\																					 
G99-21	&	$	5560	~_{-~	22	}^{+~	23	}	$	&	$	4.47 	 ~_{-~	0.04	}^{+~	0.03 	}	$	&	$	12.68 	~_{-~	2.03	 }^{+~	2.56 	 }	$	&	$	0.78 	~_{-~	0.01	}^{+~	0.01 	 }	 $	&	$	13.50 	~_{-~	1.50	}^{+~	1.00 	}	$	&	$	 0.77 	 ~_{-~	0.01	}^{+~	 0.01 	}	$	&	A	\\
G159-50	&	$	5713	~_{-~	20	}^{+~	22	}	$	&	$	4.44 	 ~_{-~	0.04	}^{+~	0.04 	}	$	&	$	14.16 	~_{-~	2.22	 }^{+~	2.00 	 }	$	&	$	0.75 	~_{-~	0.01	}^{+~	0.01 	 }	 $	&	$	15.00 	~_{-~	2.00	}^{+~	1.50 	}	$	&	$	 0.75 	 ~_{-~	0.01	}^{+~	 0.01 	}	$	&	A	\\
G180-24	&	$	6135	~_{-~	24	}^{+~	28	}	$	&	$	4.22 	 ~_{-~	0.05	}^{+~	0.03 	}	$	&	$	11.59 	~_{-~	0.61	 }^{+~	0.49 	 }	$	&	$	0.80 	~_{-~	0.01	}^{+~	0.01 	 }	 $	&	$	12.80 	~_{-~	0.30	}^{+~	0.60 	}	$	&	$	 0.79 	 ~_{-~	0.01	}^{+~	 0.01 	}	$	&	A	\\
G188-22	&	$	6121	~_{-~	24	}^{+~	19	}	$	&	$	4.20 	 ~_{-~	0.04	}^{+~	0.03 	}	$	&	$	11.04 	~_{-~	0.39	 }^{+~	0.53 	 }	$	&	$	0.82 	~_{-~	0.01	}^{+~	0.01 	 }	 $	&	$	12.50 	~_{-~	0.50	}^{+~	0.50 	}	$	&	$	 0.80 	 ~_{-~	0.00	}^{+~	 0.01 	}	$	&	A	\\
HD51754	&	$	5854	~_{-~	25	}^{+~	25	}	$	&	$	4.32 	 ~_{-~	0.02	}^{+~	0.04 	}	$	&	$	9.74 	~_{-~	1.01	 }^{+~	0.63 	 }	$	&	$	0.90 	~_{-~	0.01	}^{+~	0.00 	 }	 $	&	$	10.00 	~_{-~	1.00	}^{+~	1.00 	}	$	&	$	 0.88 	 ~_{-~	0.01	}^{+~	 0.01 	}	$	&	A	\\
HD111980	&	$	5874	~_{-~	20	}^{+~	23	}	$	&	$	 3.97 	 ~_{-~	0.02	}^{+~	0.05 	}	$	&	$	11.42 	~_{-~	 0.50	 }^{+~	 0.97 	}	$	&	$	0.84 	~_{-~	0.01	}^{+~	 0.02 	}	 $	&	$	12.50 	~_{-~	0.50	}^{+~	0.50 	}	$	 &	$	 0.84 	 ~_{-~	0.01	 }^{+~	0.00 	}	$	&	A	\\
HD113679	&	$	5765	~_{-~	22	}^{+~	20	}	$	&	$	 4.04 	 ~_{-~	0.03	}^{+~	0.03 	}	$	&	$	11.74 	~_{-~	 0.77	 }^{+~	 0.85 	}	$	&	$	0.90 	~_{-~	0.01	}^{+~	 0.01 	}	 $	&	$	12.00 	~_{-~	0.50	}^{+~	0.50 	}	$	 &	$	 0.89 	 ~_{-~	0.01	 }^{+~	0.01 	}	$	&	A	\\
HD121004	&	$	5755	~_{-~	22	}^{+~	21	}	$	&	$	 4.44 	 ~_{-~	0.04	}^{+~	0.03 	}	$	&	$	10.17 	~_{-~	 1.73	 }^{+~	 1.93 	}	$	&	$	0.82 	~_{-~	0.01	}^{+~	 0.01 	}	 $	&	$	11.50 	~_{-~	2.00	}^{+~	1.00 	}	$	 &	$	 0.81 	 ~_{-~	0.01	 }^{+~	0.01 	}	$	&	A	\\
HD132475	&	$	5758	~_{-~	25	}^{+~	15	}	$	&	$	 3.82 	 ~_{-~	0.01	}^{+~	0.01 	}	$	&	$	12.08 	~_{-~	 0.52	 }^{+~	 1.11 	}	$	&	$	0.80 	~_{-~	0.01	}^{+~	 0.01 	}	 $	&	$	11.50 	~_{-~	1.49	}^{+~	0.50 	}	$	 &	$	 0.85 	 ~_{-~	0.02	 }^{+~	0.02 	}	$	&	A	\\
HD148816	&	$	5923	~_{-~	21	}^{+~	20	}	$	&	$	 4.18 	 ~_{-~	0.04	}^{+~	0.03 	}	$	&	$	10.66 	~_{-~	 0.43	 }^{+~	 0.48 	}	$	&	$	0.89 	~_{-~	0.01	}^{+~	 0.01 	}	 $	&	$	11.50 	~_{-~	0.50	}^{+~	0.50 	}	$	 &	$	 0.87 	 ~_{-~	0.01	 }^{+~	0.01 	}	$	&	A	\\
HD159482	&	$	5827	~_{-~	27	}^{+~	26	}	$	&	$	 4.34 	 ~_{-~	0.02	}^{+~	0.04 	}	$	&	$	11.64 	~_{-~	 1.06	 }^{+~	 0.79 	}	$	&	$	0.83 	~_{-~	0.00	}^{+~	 0.01 	}	 $	&	$	12.00 	~_{-~	0.50	}^{+~	1.00 	}	$	 &	$	 0.82 	 ~_{-~	0.01	 }^{+~	0.01 	}	$	&	A	\\
HD160693	&	$	5801	~_{-~	6	}^{+~	24	}	$	&	$	 4.33 	 ~_{-~	0.03	}^{+~	0.05 	}	$	&	$	9.25 	~_{-~	 0.88	 }^{+~	 0.68 	}	$	&	$	0.92 	~_{-~	0.01	}^{+~	 0.01 	}	 $	&	$	10.00 	~_{-~	1.50	}^{+~	0.50 	}	$	 &	$	 0.89 	 ~_{-~	0.01	 }^{+~	0.01 	}	$	&	A	\\
HD179626	&	$	5957	~_{-~	23	}^{+~	22	}	$	&	$	 4.17 	 ~_{-~	0.04	}^{+~	0.04 	}	$	&	$	12.71 	~_{-~	 0.53	 }^{+~	 0.62 	}	$	&	$	0.82 	~_{-~	0.02	}^{+~	 0.01 	}	 $	&	$	13.50 	~_{-~	0.50	}^{+~	0.50 	}	$	 &	$	 0.81 	 ~_{-~	0.01	 }^{+~	0.01 	}	$	&	A	\\
HD222766	&	$	5423	~_{-~	16	}^{+~	27	}	$	&	$	 4.37 	 ~_{-~	0.02	}^{+~	0.03 	}	$	&	$	22.19 	~_{-~	 1.16	 }^{+~	 1.17 	}	$	&	$	0.72 	~_{-~	0.01	}^{+~	 0.01 	}	 $	&	$	22.00 	~_{-~	1.00	}^{+~	0.50 	}	$	 &	$	 0.71 	 ~_{-~	0.01	 }^{+~	0.01 	}	$	&	A	\\
HD230409	&	$	5388	~_{-~	24	}^{+~	20	}	$	&	$	 4.62 	 ~_{-~	0.04	}^{+~	0.03 	}	$	&	$	6.16 	~_{-~	 3.74	 }^{+~	 4.33 	}	$	&	$	0.73 	~_{-~	0.01	}^{+~	 0.01 	}	 $	&	$	12.01 	~_{-~	4.50	}^{+~	4.00 	}	$	 &	$	 0.70 	 ~_{-~	0.01	 }^{+~	0.02 	}	$	&	A	\\
\noalign{\smallskip}\hline\noalign{\smallskip}
CD-453283	&	$	5686	~_{-~	23	}^{+~	20	}	$	&	$	 4.61 	 ~_{-~	0.04	}^{+~	0.03 	}	$	&	$	4.51 	~_{-~	 2.82	 }^{+~	 3.36 	}	$	&	$	0.75 	~_{-~	0.01	}^{+~	 0.02 	}	 $	&	$	5.50 	~_{-~	3.00	}^{+~	3.50 	}	$	 &	$	 0.76 	 ~_{-~	0.02	 }^{+~	0.01 	}	$	&	B	\\
CD-514628	&	$	6299	~_{-~	23	}^{+~	24	}	$	&	$	 4.27 	 ~_{-~	0.03	}^{+~	0.04 	}	$	&	$	8.56 	~_{-~	 0.57	 }^{+~	 0.56 	}	$	&	$	0.84 	~_{-~	0.01	}^{+~	 0.02 	}	 $	&	$	9.50 	~_{-~	0.50	}^{+~	0.50 	}	$	 &	$	 0.84 	 ~_{-~	0.00	 }^{+~	0.01 	}	$	&	B	\\
CD-571633	&	$	5979	~_{-~	26	}^{+~	25	}	$	&	$	 4.29 	 ~_{-~	0.02	}^{+~	0.03 	}	$	&	$	12.62 	~_{-~	 0.62	 }^{+~	 0.65 	}	$	&	$	0.80 	~_{-~	0.01	}^{+~	 0.01 	}	 $	&	$	13.00 	~_{-~	0.80	}^{+~	0.60 	}	$	 &	$	 0.80 	 ~_{-~	0.01	 }^{+~	0.01 	}	$	&	B	\\
CD-610282	&	$	5860	~_{-~	3	}^{+~	27	}	$	&	$	 4.33 	 ~_{-~	0.02	}^{+~	0.03 	}	$	&	$	18.14 	~_{-~	 0.69	 }^{+~	 0.73 	}	$	&	$	0.70 	~_{-~	0.00	}^{+~	 0.01 	}	 $	&	$	18.00 	~_{-~	1.00	}^{+~	1.00 	}	$	 &	$	 0.72 	 ~_{-~	0.01	 }^{+~	0.01 	}	$	&	B	\\
G05-19	&	$	5775	~_{-~	15	}^{+~	17	}	$	&	$	4.29 	 ~_{-~	0.04	}^{+~	0.04 	}	$	&	$	22.51 	~_{-~	0.95	 }^{+~	0.91 	 }	$	&	$	0.67 	~_{-~	0.00	}^{+~	0.01 	 }	 $	&	$	22.00 	~_{-~	1.00	}^{+~	0.50 	}	$	&	$	 0.69 	 ~_{-~	0.01	}^{+~	 0.01 	}	$	&	B	\\
G20-15	&	$	6156	~_{-~	21	}^{+~	25	}	$	&	$	4.35 	 ~_{-~	0.02	}^{+~	0.02 	}	$	&	$	11.24 	~_{-~	0.79	 }^{+~	0.77 	 }	$	&	$	0.76 	~_{-~	0.00	}^{+~	0.01 	 }	 $	&	$	12.00 	~_{-~	1.00	}^{+~	0.50 	}	$	&	$	 0.78 	 ~_{-~	0.01	}^{+~	 0.01 	}	$	&	B	\\
G46-31	&	$	6026	~_{-~	27	}^{+~	22	}	$	&	$	4.29 	 ~_{-~	0.03	}^{+~	0.02 	}	$	&	$	10.31 	~_{-~	0.44	 }^{+~	0.70 	 }	$	&	$	0.84 	~_{-~	0.01	}^{+~	0.02 	 }	 $	&	$	10.50 	~_{-~	0.50	}^{+~	0.50 	}	$	&	$	 0.85 	 ~_{-~	0.01	}^{+~	 0.01 	}	$	&	B	\\
G53-41	&	$	5977	~_{-~	27	}^{+~	27	}	$	&	$	4.29 	 ~_{-~	0.03	}^{+~	0.03 	}	$	&	$	15.51 	~_{-~	0.77	 }^{+~	0.79 	 }	$	&	$	0.73 	~_{-~	0.00	}^{+~	0.01 	 }	 $	&	$	15.00 	~_{-~	1.00	}^{+~	0.50 	}	$	&	$	 0.76 	 ~_{-~	0.01	}^{+~	 0.01 	}	$	&	B	\\
G56-36	&	$	6062	~_{-~	24	}^{+~	24	}	$	&	$	4.31 	 ~_{-~	0.02	}^{+~	0.04 	}	$	&	$	10.13 	~_{-~	0.63	 }^{+~	0.62 	 }	$	&	$	0.83 	~_{-~	0.01	}^{+~	0.01 	 }	 $	&	$	9.50 	~_{-~	1.00	}^{+~	0.50 	}	$	&	$	 0.85 	 ~_{-~	0.01	}^{+~	 0.01 	}	$	&	B	\\
G66-22	&	$	5296	~_{-~	21	}^{+~	24	}	$	&	$	4.43 	 ~_{-~	0.03	}^{+~	0.06 	}	$	&	$	30.85 	~_{-~	4.79	 }^{+~	2.75 	 }	$	&	$	0.62 	~_{-~	0.00	}^{+~	0.01 	 }	 $	&	$	29.50 	~_{-~	2.00	}^{+~	2.00 	}	$	&	$	 0.61 	 ~_{-~	0.02	}^{+~	 0.01 	}	$	&	B	\\
G75-31	&	$	6136	~_{-~	21	}^{+~	21	}	$	&	$	4.03 	 ~_{-~	0.04	}^{+~	0.03 	}	$	&	$	8.73 	~_{-~	0.56	 }^{+~	0.42 	 }	$	&	$	0.91 	~_{-~	0.02	}^{+~	0.01 	 }	 $	&	$	9.00 	~_{-~	0.00	}^{+~	0.50 	}	$	&	$	 0.90 	 ~_{-~	0.01	}^{+~	 0.01 	}	$	&	B	\\
G82-05	&	$	5338	~_{-~	22	}^{+~	23	}	$	&	$	4.52 	 ~_{-~	0.05	}^{+~	0.03 	}	$	&	$	19.03 	~_{-~	3.66	 }^{+~	4.26 	 }	$	&	$	0.68 	~_{-~	0.02	}^{+~	0.01 	 }	 $	&	$	22.00 	~_{-~	3.00	}^{+~	3.00 	}	$	&	$	 0.66 	 ~_{-~	0.02	}^{+~	 0.01 	}	$	&	B	\\
G112-43	&	$	6207	~_{-~	21	}^{+~	21	}	$	&	$	4.03 	 ~_{-~	0.04	}^{+~	0.03 	}	$	&	$	8.45 	~_{-~	0.46	 }^{+~	0.90 	 }	$	&	$	0.90 	~_{-~	0.03	}^{+~	0.01 	 }	 $	&	$	9.50 	~_{-~	0.00	}^{+~	0.50 	}	$	&	$	 0.88 	 ~_{-~	0.01	}^{+~	 0.01 	}	$	&	B	\\
G119-64	&	$	6342	~_{-~	20	}^{+~	14	}	$	&	$	4.13 	 ~_{-~	0.03	}^{+~	0.04 	}	$	&	$	8.40 	~_{-~	0.47	 }^{+~	0.28 	 }	$	&	$	0.87 	~_{-~	0.00	}^{+~	0.01 	 }	 $	&	$	10.50 	~_{-~	0.00	}^{+~	0.50 	}	$	&	$	 0.84 	 ~_{-~	0.01	}^{+~	 0.01 	}	$	&	B	\\
G150-40	&	$	6079	~_{-~	22	}^{+~	22	}	$	&	$	4.12 	 ~_{-~	0.04	}^{+~	0.03 	}	$	&	$	9.37 	~_{-~	0.39	 }^{+~	0.46 	 }	$	&	$	0.90 	~_{-~	0.02	}^{+~	0.01 	 }	 $	&	$	9.00 	~_{-~	0.00	}^{+~	0.50 	}	$	&	$	 0.91 	 ~_{-~	0.00	}^{+~	 0.01 	}	$	&	B	\\
G170-56	&	$	6111	~_{-~	22	}^{+~	23	}	$	&	$	4.12 	 ~_{-~	0.04	}^{+~	0.03 	}	$	&	$	9.68 	~_{-~	0.42	 }^{+~	0.48 	 }	$	&	$	0.88 	~_{-~	0.02	}^{+~	0.01 	 }	 $	&	$	9.50 	~_{-~	0.50	}^{+~	0.50 	}	$	&	$	 0.88 	 ~_{-~	0.01	}^{+~	 0.02 	}	$	&	B	\\
HD3567	&	$	6178	~_{-~	20	}^{+~	24	}	$	&	$	4.02 	 ~_{-~	0.04	}^{+~	0.03 	}	$	&	$	8.73 	~_{-~	0.83	 }^{+~	0.40 	 }	$	&	$	0.90 	~_{-~	0.02	}^{+~	0.03 	 }	 $	&	$	9.50 	~_{-~	0.50	}^{+~	0.50 	}	$	&	$	 0.90 	 ~_{-~	0.01	}^{+~	 0.00 	}	$	&	B	\\
HD59392	&	$	6127	~_{-~	22	}^{+~	36	}	$	&	$	3.88 	 ~_{-~	0.03	}^{+~	0.04 	}	$	&	$	8.86 	~_{-~	0.93	 }^{+~	1.40 	 }	$	&	$	0.87 	~_{-~	0.04	}^{+~	0.03 	 }	 $	&	$	10.00 	~_{-~	0.50	}^{+~	1.00 	}	$	&	$	 0.86 	 ~_{-~	0.01	}^{+~	 0.01 	}	$	&	B	\\
HD103723	&	$	6047	~_{-~	19	}^{+~	23	}	$	&	$	 4.22 	 ~_{-~	0.04	}^{+~	0.03 	}	$	&	$	10.27 	~_{-~	 0.45	 }^{+~	 0.49 	}	$	&	$	0.86 	~_{-~	0.01	}^{+~	 0.01 	}	 $	&	$	10.00 	~_{-~	0.20	}^{+~	1.00 	}	$	 &	$	 0.87 	 ~_{-~	0.01	 }^{+~	0.01 	}	$	&	B	\\
HD105004	&	$	5858	~_{-~	27	}^{+~	25	}	$	&	$	 4.33 	 ~_{-~	0.02	}^{+~	0.03 	}	$	&	$	14.00 	~_{-~	 0.65	 }^{+~	 0.64 	}	$	&	$	0.78 	~_{-~	0.01	}^{+~	 0.01 	}	 $	&	$	13.50 	~_{-~	1.00	}^{+~	1.00 	}	$	 &	$	 0.79 	 ~_{-~	0.01	 }^{+~	0.01 	}	$	&	B	\\
HD163810	&	$	5591	~_{-~	22	}^{+~	23	}	$	&	$	 4.60 	 ~_{-~	0.03	}^{+~	0.02 	}	$	&	$	11.69 	~_{-~	 2.42	 }^{+~	 2.89 	}	$	&	$	0.68 	~_{-~	0.02	}^{+~	 0.01 	}	 $	&	$	12.50 	~_{-~	3.50	}^{+~	3.00 	}	$	 &	$	 0.68 	 ~_{-~	0.01	 }^{+~	0.01 	}	$	&	B	\\
HD193901	&	$	5742	~_{-~	15	}^{+~	23	}	$	&	$	 4.40 	 ~_{-~	0.03	}^{+~	0.05 	}	$	&	$	19.32 	~_{-~	 1.95	 }^{+~	 1.03 	}	$	&	$	0.69 	~_{-~	0.01	}^{+~	 0.01 	}	 $	&	$	18.00 	~_{-~	1.50	}^{+~	1.00 	}	$	 &	$	 0.70 	 ~_{-~	0.01	 }^{+~	0.01 	}	$	&	B	\\
HD194598	&	$	6058	~_{-~	24	}^{+~	24	}	$	&	$	 4.30 	 ~_{-~	0.02	}^{+~	0.04 	}	$	&	$	11.81 	~_{-~	 0.69	 }^{+~	 0.64 	}	$	&	$	0.79 	~_{-~	0.01	}^{+~	 0.01 	}	 $	&	$	11.00 	~_{-~	0.50	}^{+~	1.00 	}	$	 &	$	 0.80 	 ~_{-~	0.01	 }^{+~	0.01 	}	$	&	B	\\
HD219617	&	$	5988	~_{-~	19	}^{+~	11	}	$	&	$	 4.30 	 ~_{-~	0.04	}^{+~	0.02 	}	$	&	$	17.46 	~_{-~	 0.69	 }^{+~	 0.56 	}	$	&	$	0.70 	~_{-~	0.00	}^{+~	 0.01 	}	 $	&	$	17.50 	~_{-~	1.00	}^{+~	0.50 	}	$	 &	$	 0.72 	 ~_{-~	0.01	 }^{+~	0.01 	}	$	&	B	\\
HD284248	&	$	6257	~_{-~	6	}^{+~	28	}	$	&	$	 4.24 	 ~_{-~	0.02	}^{+~	0.02 	}	$	&	$	10.56 	~_{-~	 0.42	 }^{+~	 0.32 	}	$	&	$	0.80 	~_{-~	0.00	}^{+~	 0.01 	}	 $	&	$	12.00 	~_{-~	0.50	}^{+~	0.50 	}	$	 &	$	 0.80 	 ~_{-~	0.01	 }^{+~	0.01 	}	$	&	B	\\																																																									 \noalign{\smallskip}\hline\noalign{\smallskip}
BD-213420	&	$	5914	~_{-~	26	}^{+~	25	}	$	&	$	 4.31 	 ~_{-~	0.03	}^{+~	0.03 	}	$	&	$	14.48 	~_{-~	 0.70	 }^{+~	 0.74 	}	$	&	$	0.76 	~_{-~	0.01	}^{+~	 0.01 	}	 $	&	$	15.50 	~_{-~	1.00	}^{+~	0.50 	}	$	 &	$	 0.76 	 ~_{-~	0.01	 }^{+~	0.01 	}	$	&	C	\\
CD-333337	&	$	6111	~_{-~	18	}^{+~	18	}	$	&	$	 3.88 	 ~_{-~	0.03	}^{+~	0.03 	}	$	&	$	7.83 	~_{-~	 0.80	 }^{+~	 0.61 	}	$	&	$	0.92 	~_{-~	0.02	}^{+~	 0.02 	}	 $	&	$	8.50 	~_{-~	0.50	}^{+~	1.00 	}	$	 &	$	 0.91 	 ~_{-~	0.02	 }^{+~	0.02 	}	$	&	C	\\
HD4308	&	$	5699	~_{-~	9	}^{+~	19	}	$	&	$	4.34 	 ~_{-~	0.02	}^{+~	0.04 	}	$	&	$	10.82 	~_{-~	1.11	 }^{+~	0.52 	 }	$	&	$	0.89 	~_{-~	0.00	}^{+~	0.01 	 }	 $	&	$	10.00 	~_{-~	1.00	}^{+~	0.50 	}	$	&	$	 0.91 	 ~_{-~	0.01	}^{+~	 0.01 	}	$	&	C	\\
HD17820	&	$	5868	~_{-~	25	}^{+~	24	}	$	&	$	4.29 	 ~_{-~	0.04	}^{+~	0.02 	}	$	&	$	10.97 	~_{-~	0.50	 }^{+~	0.54 	 }	$	&	$	0.87 	~_{-~	0.01	}^{+~	0.01 	 }	 $	&	$	12.00 	~_{-~	0.50	}^{+~	0.50 	}	$	&	$	 0.85 	 ~_{-~	0.01	}^{+~	 0.01 	}	$	&	C	\\
HD22879	&	$	5857	~_{-~	26	}^{+~	25	}	$	&	$	4.30 	 ~_{-~	0.03	}^{+~	0.02 	}	$	&	$	12.86 	~_{-~	0.65	 }^{+~	0.66 	 }	$	&	$	0.82 	~_{-~	0.01	}^{+~	0.01 	 }	 $	&	$	13.50 	~_{-~	0.50	}^{+~	1.00 	}	$	&	$	 0.80 	 ~_{-~	0.01	}^{+~	 0.01 	}	$	&	C	\\
HD25704	&	$	5980	~_{-~	23	}^{+~	24	}	$	&	$	4.30 	 ~_{-~	0.02	}^{+~	0.03 	}	$	&	$	11.24 	~_{-~	0.59	 }^{+~	0.59 	 }	$	&	$	0.82 	~_{-~	0.00	}^{+~	0.01 	 }	 $	&	$	11.50 	~_{-~	1.00	}^{+~	0.50 	}	$	&	$	 0.83 	 ~_{-~	0.01	}^{+~	 0.01 	}	$	&	C	\\
HD65907	&	$	5997	~_{-~	16	}^{+~	20	}	$	&	$	4.41 	 ~_{-~	0.03	}^{+~	0.03 	}	$	&	$	3.74 	~_{-~	0.88	 }^{+~	0.86 	 }	$	&	$	1.01 	~_{-~	0.01	}^{+~	0.01 	 }	 $	&	$	4.00 	~_{-~	1.00	}^{+~	0.50 	}	$	&	$	 1.02 	 ~_{-~	0.01	}^{+~	 0.01 	}	$	&	C	\\
HD76932	&	$	5978	~_{-~	22	}^{+~	21	}	$	&	$	4.18 	 ~_{-~	0.04	}^{+~	0.03 	}	$	&	$	11.53 	~_{-~	0.50	 }^{+~	0.55 	 }	$	&	$	0.84 	~_{-~	0.01	}^{+~	0.01 	 }	 $	&	$	11.50 	~_{-~	0.50	}^{+~	0.50 	}	$	&	$	 0.86 	 ~_{-~	0.01	}^{+~	 0.01 	}	$	&	C	\\
HD77110	&	$	5739	~_{-~	20	}^{+~	16	}	$	&	$	4.39 	 ~_{-~	0.03	}^{+~	0.03 	}	$	&	$	9.77 	~_{-~	1.16	 }^{+~	1.34 	 }	$	&	$	0.88 	~_{-~	0.01	}^{+~	0.01 	 }	 $	&	$	10.50 	~_{-~	1.50	}^{+~	1.00 	}	$	&	$	 0.86 	 ~_{-~	0.01	}^{+~	 0.01 	}	$	&	C	\\
HD97320	&	$	6135	~_{-~	21	}^{+~	23	}	$	&	$	4.21 	 ~_{-~	0.04	}^{+~	0.03 	}	$	&	$	9.95 	~_{-~	0.42	 }^{+~	0.41 	 }	$	&	$	0.85 	~_{-~	0.01	}^{+~	0.01 	 }	 $	&	$	11.00 	~_{-~	0.50	}^{+~	0.50 	}	$	&	$	 0.83 	 ~_{-~	0.01	}^{+~	 0.01 	}	$	&	C	\\
HD106516	&	$	6325	~_{-~	20	}^{+~	22	}	$	&	$	 4.43 	 ~_{-~	0.04	}^{+~	0.03 	}	$	&	$	1.49 	~_{-~	 0.87	 }^{+~	 0.84 	}	$	&	$	1.03 	~_{-~	0.01	}^{+~	 0.01 	}	 $	&	$	2.00 	~_{-~	0.50	}^{+~	1.50 	}	$	 &	$	 1.01 	 ~_{-~	0.01	 }^{+~	0.00 	}	$	&	C	\\
HD111232	&	$	5543	~_{-~	18	}^{+~	19	}	$	&	$	 4.43 	 ~_{-~	0.04	}^{+~	0.03 	}	$	&	$	11.62 	~_{-~	 -1.34 	 }^{+~	 1.79 	}	$	&	$	0.83 	~_{-~	0.01	}^{+~	 0.01 	}	 $	&	$	13.00 	~_{-~	2.00	}^{+~	1.00 	}	$	 &	$	 0.82 	 ~_{-~	0.01	 }^{+~	0.01 	}	$	&	C	\\																						 
HD114762	&	$	5961	~_{-~	23	}^{+~	22	}	$	&	$	 4.25 	 ~_{-~	0.04	}^{+~	0.03 	}	$	&	$	9.98 	~_{-~	 0.47	 }^{+~	 0.47 	}	$	&	$	0.89 	~_{-~	0.01	}^{+~	 0.02 	}	 $	&	$	11.00 	~_{-~	0.50	}^{+~	0.50 	}	$	 &	$	 0.86 	 ~_{-~	0.01	 }^{+~	0.01 	}	$	&	C	\\
HD120559	&	$	5486	~_{-~	22	}^{+~	23	}	$	&	$	 4.59 	 ~_{-~	0.04	}^{+~	0.03 	}	$	&	$	8.29 	~_{-~	 3.73	 }^{+~	 3.63 	}	$	&	$	0.74 	~_{-~	0.16	}^{+~	 0.01 	}	 $	&	$	12.51 	~_{-~	3.50	}^{+~	4.50 	}	$	 &	$	 0.71 	 ~_{-~	0.01	 }^{+~	0.01 	}	$	&	C	\\
HD126681	&	$	5595	~_{-~	22	}^{+~	22	}	$	&	$	 4.51 	 ~_{-~	0.05	}^{+~	0.03 	}	$	&	$	17.33 	~_{-~	 3.01	 }^{+~	 3.05 	}	$	&	$	0.68 	~_{-~	0.22	}^{+~	 0.01 	}	 $	&	$	20.01 	~_{-~	2.50	}^{+~	3.00 	}	$	 &	$	 0.67 	 ~_{-~	0.01	 }^{+~	0.01 	}	$	&	C	\\
HD136352	&	$	5722	~_{-~	4	}^{+~	24	}	$	&	$	 4.34 	 ~_{-~	0.02	}^{+~	0.05 	}	$	&	$	9.65 	~_{-~	 0.80	 }^{+~	 0.52 	}	$	&	$	0.92 	~_{-~	0.01	}^{+~	 0.01 	}	 $	&	$	10.00 	~_{-~	1.00	}^{+~	0.50 	}	$	 &	$	 0.90 	 ~_{-~	0.01	 }^{+~	0.01 	}	$	&	C	\\
HD175179	&	$	5805	~_{-~	20	}^{+~	19	}	$	&	$	 4.40 	 ~_{-~	0.04	}^{+~	0.04 	}	$	&	$	9.87 	~_{-~	 1.61	 }^{+~	 1.51 	}	$	&	$	0.85 	~_{-~	0.01	}^{+~	 0.01 	}	 $	&	$	11.50 	~_{-~	1.49	}^{+~	1.00 	}	$	 &	$	 0.83 	 ~_{-~	0.01	 }^{+~	0.01 	}	$	&	C	\\
HD189558	&	$	5702	~_{-~	19	}^{+~	23	}	$	&	$	 3.84 	 ~_{-~	0.03	}^{+~	0.02 	}	$	&	$	11.13 	~_{-~	 1.28	 }^{+~	 0.51 	}	$	&	$	0.85 	~_{-~	0.02	}^{+~	 0.02 	}	 $	&	$	11.00 	~_{-~	1.00	}^{+~	1.00 	}	$	 &	$	 0.88 	 ~_{-~	0.03	 }^{+~	0.03 	}	$	&	C	\\
HD199289	&	$	5920	~_{-~	15	}^{+~	12	}	$	&	$	 4.31 	 ~_{-~	0.03	}^{+~	0.03 	}	$	&	$	13.64 	~_{-~	 0.50	 }^{+~	 0.53 	}	$	&	$	0.78 	~_{-~	0.00	}^{+~	 0.01 	}	 $	&	$	14.00 	~_{-~	0.50	}^{+~	0.50 	}	$	 &	$	 0.78 	 ~_{-~	0.00	 }^{+~	0.01 	}	$	&	C	\\
HD205650	&	$	5794	~_{-~	28	}^{+~	28	}	$	&	$	 4.33 	 ~_{-~	0.03	}^{+~	0.03 	}	$	&	$	18.43 	~_{-~	 1.18	 }^{+~	 1.01 	}	$	&	$	0.71 	~_{-~	0.19	}^{+~	 0.01 	}	 $	&	$	19.00 	~_{-~	1.50	}^{+~	1.00 	}	$	 &	$	 0.71 	 ~_{-~	0.01	 }^{+~	0.01 	}	$	&	C	\\
HD241253	&	$	5941	~_{-~	26	}^{+~	26	}	$	&	$	 4.32 	 ~_{-~	0.02	}^{+~	0.03 	}	$	&	$	13.40 	~_{-~	 0.77	 }^{+~	 0.76 	}	$	&	$	0.77 	~_{-~	0.00	}^{+~	 0.01 	}	 $	&	$	13.50 	~_{-~	1.00	}^{+~	1.00 	}	$	 &	$	 0.77 	 ~_{-~	0.01	 }^{+~	0.01 	}	$	&	C	\\
\enddata
\tablenotetext{}{Population classification: A, high-$\alpha$ halo; B, low-$\alpha$ halo; C, thick-disk population.}
\end{deluxetable}

\clearpage

\appendix

\section{}
\subsection{Constructing Metal mixtures}

We fix the abundance of Fe and increase or decrease carbon, oxygen, and $\alpha$-elements to construct metal mixtures. We add the enhancement factors to the solar log $N$ values. With newly constructed log $N$, we could calculate the number fraction and the mass fraction for each element, and to construct opacity tables for the stellar models. The chemical composition in stellar models is identical to the metal mixture used in opacity tables.
The complete version of one metal mixture, [C/Fe] = 0.3, [O/Fe] = 0.7, [$\alpha$/Fe] = 0.3, is presented in Table \ref{tab:exg}. From the number fraction and mass fraction we find that C and O take larger part than the $\alpha$-elements, either in a scaled-solar metal-mixture or in the CO extreme mix. The high-$\alpha$ metal mixtures and low-$\alpha$ metal mixtures are presented in Table \ref{tab:higha} and \ref{tab:lowA}.

\subsection[]{Comparison Results from a High-$\alpha$ and Low-$\alpha$ Metal Mixture with Similar C and O Abundances}

From observed $\alpha$-elements abundances, we note that some low-$\alpha$ population stars in fact have high $\alpha$-elements abundances. For HD 59392, HD 219617, and G119-64, they have [$\alpha$/Fe] $\sim$ 0.3. We made a test using different metal mixtures with similar [C/Fe] and [O/Fe], but with different [$\alpha$/Fe] to calculate these stars to determine the influence of the $\alpha$-elements. The metal mixtures are presented in Table \ref{tab:comp13}.

From the information in Table \ref{tab:comp13} we know that with similar C and O but different $\alpha$-enhancement, the overall number fraction and mass fraction are slightly modified. The main parts of the metal mixture, i.e., the C, N, and O abundances, are not obviously different. We present the relationship between [Fe/H] and $Z$ in the different metal mixtures in Table \ref{tab:compZ} and notice that with similar C and O, the difference of the $Z$ values from different $\alpha$-enhancements is very small for the metal-poor stars.

The observation constraints and the modeling results are presented in Table \ref{tab:compresult}. The differences of the estimated ages are typically $<$ 0.5 Gyr, which is much smaller than the estimation error, but for HD 219617 the difference would be 1.3 Gyr, which is near to the estimation uncertainty. From all the comparison results we can say that C and O are the main factors that influence the stellar models, especially for metal-poor stars. Thus we make an approximation of [$\alpha$/Fe] = 0.1 for all the low-$\alpha$ halo stars.

\clearpage

\begin{deluxetable}{ccccccccccccc}
\tablecaption{metal mixture for [O/Fe] = 0.7, [C/Fe] = 0.3, [$\alpha$/Fe] = 0.3. \label{tab:exg}}
\tablehead{
Element &  Weight  & $\log$ $N_\odot$ & N.F.$_\odot$  & M.F.$_\odot$     &	 Enhancement  &    $\log$ $N$      & N.F. & M.F.
}
\startdata
  	C	 	&	 	12.0107   	&	   	 8.52 	  	&  	 0.245197 	 	 &	 	 0.171014 	 	 &	 	 0.30 	 &	  	8.82 	 &	 	 0.143353   	&	 	 0.104808   \\
	N		&		14.0067 		&		7.92 		 &		 0.061591 		 &		 0.050094 		&	...			 &		7.92 		 &		 0.018047 		 &		 0.015387 		 \\
 O	  	&	 15.9994 	 	&	 	 8.83 	  	&	  	 0.500628 	 	 &	 	 0.465111  	 &	  	 0.70 	  	&	   	9.53 	 	&	  	 0.735202 	  	 &	 	 0.716008 	 	 \\
	F		&		18.9984 		&		4.56 		 &		 0.000027 		 &		 0.000030 		&		...		 &		4.56 		 &		 0.000008 		 &		 0.000009 		 \\
	Ne		&		20.1797 		&		8.08 		 &		 0.089026 		 &		 0.104316 		&		 0.30 		&		 8.38 		&		 0.052048 		 &		 0.063931 		\\
	Na		&		22.9898 		&		6.33 		 &		 0.001583 		 &		 0.002113 		&		...		 &		6.33 		 &		 0.000464 		 &		 0.000649 		 \\
	Mg		&		24.3050 		&		7.58 		 &		 0.028152 		 &		 0.039733 		&		 0.30 		&		 7.88 		&		 0.016459 		 &		 0.024351 		\\
	Al		&		26.9815 		&		6.47 		 &		 0.002185 		 &		 0.003424 		&		...		 &		6.47 		 &		 0.000640 		 &		 0.001052 		 \\
	Si		&		28.0855 		&		7.55 		 &		 0.026273 		 &		 0.042848 		&		 0.30 		&		 7.85 		&		 0.015361 		 &		 0.026260 		\\
	P		&		30.9738 		&		5.45 		 &		 0.000209 		 &		 0.000375 		&		...		 &		5.45 		 &		 0.000061 		 &		 0.000115 		 \\
	S		&		32.0650 		&		7.33 		 &		 0.015831 		 &		 0.029472 		&		 0.30 		&		 7.63 		&		 0.009256 		 &		 0.018062 		\\
	Cl		&		35.4530 		&		5.50 		 &		 0.000234 		 &		 0.000482 		&		...		 &		5.50 		 &		 0.000069 		 &		 0.000148 		 \\
	Ar		&		39.9480 		&		6.40 		 &		 0.001860 		 &		 0.004315 		&		...		 &		6.40 		 &		 0.000545 		 &		 0.001325 		 \\
	K		&		39.0983 		&		5.12 		 &		 0.000098 		 &		 0.000222 		&		...		 &		5.12 		 &		 0.000029 		 &		 0.000068 		 \\
	Ca		&		40.0780 		&		6.36 		 &		 0.001696 		 &		 0.003948 		&		 0.30 		&		 6.66 		&		 0.000992 		 &		 0.002420 		\\
	Sc		&		44.9559 		&		3.17 		 &		 0.000001 		 &		 0.000003 		&		...		 &		3.17 		 &		 0.000000 		 &		 0.000001 		 \\
	Ti		&		47.8670 		&		5.02 		 &		 0.000078 		 &		 0.000216 		&		 0.30 		&		 5.32 		&		 0.000045 		 &		 0.000132 		\\
	V		&		50.9415 		&		4.00 		 &		 0.000007 		 &		 0.000022 		&		...		 &		4.00 		 &		 0.000002 		 &		 0.000007 		 \\
	Cr		&		51.9961 		&		5.67 		 &		 0.000346 		 &		 0.001046 		&		...		 &		5.67 		 &		 0.000101 		 &		 0.000321 		 \\
	Mn		&		54.9380 		&		5.39 		 &		 0.000182 		 &		 0.000580 		&		...		 &		5.39 		 &		 0.000053 		 &		 0.000178 		 \\
	Fe		&		55.8450 		&		7.50 		 &		 0.023416 		 &		 0.075937 		&		...		 &		7.50 		 &		 0.006861 		 &		 0.023325 		 \\
	Co		&		58.9332 		&		4.92 		 &		 0.000062 		 &		 0.000211 		&		...		 &		4.92 		 &		 0.000018 		 &		 0.000065 		 \\
	Ni		&		58.6934 		&		6.25 		 &		 0.001317 		 &		 0.004488 		&		...		 &		6.25 		 &		 0.000386 		 &		 0.001379 		 \\
\enddata
\tablenotetext{}{The abbreviation N.F. represents the number fraction of each metal element, and M.F. means the mass fraction of each element.}
\end{deluxetable}

\clearpage

\begin{deluxetable}{ccc|cc|cc|cc|cc}
\tablecaption{Metal mixtures for high-$\alpha$ populations. \label{tab:higha}}
\tablehead{
element &  Weight  & $\log$ N$_\odot$    &	enhance  &    $\log$ N   &	 enhance  &    $\log$ N  &	 enhance  &    $\log$ N  &	 enhance  &    $\log$ N
}
\startdata
C	&	12.0107  	&	8.52 	&	0.20 	&	8.72 	 &	0.20 	 &	 8.72 	&	 0.20 	&	 8.72 	&	 0.20 	 &	8.72 	\\
N	&	14.0067  	&	7.92 	&		&	7.92 	&		 &	7.92 	 &		 &	 7.92 	&		 &	 7.92 	 \\
O	&	15.9994  	&	8.83 	&	0.60 	&	9.43 	 &	0.50 	 &	 9.33 	&	 0.40 	&	 9.23 	&	 0.30 	 &	9.13 	\\
F	&	18.9984 	&	4.56 	&		&	4.56 	&		 &	4.56 	 &		 &	 4.56 	&		 &	 4.56 	 \\
Ne	&	20.1797 	&	8.08 	&	0.30 	&	8.38 	 &	0.30 	 &	 8.38 	&	 0.30 	&	 8.38 	&	 0.20 	 &	8.28 	\\
Na	&	22.9898 	&	6.33 	&		&	6.33 	&		 &	6.33 	 &		 &	 6.33 	&		 &	 6.33 	 \\
Mg	&	24.3050 	&	7.58 	&	0.30 	&	7.88 	 &	0.30 	 &	 7.88 	&	 0.30 	&	 7.88 	&	 0.20 	 &	7.78 	\\
Al	&	26.9815 	&	6.47 	&		&	6.47 	&		 &	6.47 	 &		 &	 6.47 	&		 &	 6.47 	 \\
Si	&	28.0855 	&	7.55 	&	0.30 	&	7.85 	 &	0.30 	 &	 7.85 	&	 0.30 	&	 7.85 	&	 0.20 	 &	7.75 	\\
P	&	30.9738 	&	5.45 	&		&	5.45 	&		 &	5.45 	 &		 &	 5.45 	&		 &	 5.45 	 \\
S	&	32.0650 	&	7.33 	&	0.30 	&	7.63 	 &	0.30 	 &	 7.63 	&	 0.30 	&	 7.63 	&	 0.20 	 &	7.53 	\\
Cl	&	35.4530 	&	5.50 	&		&	5.50 	&		 &	5.50 	 &		 &	 5.50 	&		 &	 5.50 	 \\
Ar	&	39.9480 	&	6.40 	&	0.00 	&	6.40 	 &		&	 6.40 	 &		 &	6.40 	 &		&	 6.40 	 \\
K	&	39.0983 	&	5.12 	&		&	5.12 	&		 &	5.12 	 &		 &	 5.12 	&		 &	 5.12 	 \\
Ca	&	40.0780 	&	6.36 	&	0.30 	&	6.66 	 &	0.30 	 &	 6.66 	&	 0.30 	&	 6.66 	&	 0.20 	 &	6.56 	\\
Sc	&	44.9559 	&	3.17 	&		&	3.17 	&		 &	3.17 	 &		 &	 3.17 	&		 &	 3.17 	 \\
Ti	&	47.8670 	&	5.02 	&	0.30 	&	5.32 	 &	0.30 	 &	 5.32 	&	 0.30 	&	 5.32 	&	 0.20 	 &	5.22 	\\
V	&	50.9415 	&	4.00 	&		&	4.00 	&		 &	4.00 	 &		 &	 4.00 	&		 &	 4.00 	 \\
Cr	&	51.9961 	&	5.67 	&		&	5.67 	&		 &	5.67 	 &		 &	 5.67 	&		 &	 5.67 	 \\
Mn	&	54.9380 	&	5.39 	&		&	5.39 	&		 &	5.39 	 &		 &	 5.39 	&		 &	 5.39 	 \\
Fe	&	55.8450 	&	7.50 	&		&	7.50 	&		 &	7.50 	 &		 &	 7.50 	&		 &	 7.50 	 \\
Co	&	58.9332 	&	4.92 	&		&	4.92 	&		 &	4.92 	 &		 &	 4.92 	&		 &	 4.92 	 \\
Ni	&	58.6934 	&	6.25 	&		&	6.25 	&		 &	6.25 	 &		 &	 6.25 	&		 &	 6.25 	 \\
\enddata
\end{deluxetable}

\clearpage
\begin{deluxetable}{ccc|cc|cc|cc }
\tablecaption{Metal mixtures for low-$\alpha$ halo. \label{tab:lowA}}
\tablehead{
element &  Weight  & $\log$ N$_\odot$    &	enhance  &    $\log$ N   &	 enhance  &    $\log$ N  &	 enhance  &    $\log$ N
}
\startdata
C	&	12.0107 	&	8.52 	&	0.1	&	8.62 	&	 0	&	8.52 	 &	 -0.1	 &	8.42 	 \\
N	&	14.0067 	&	7.92 	&		&	7.92 	&		 &	7.92 	 &		 &	 7.92 	\\
O	&	15.9994 	&	8.83 	&	0.5	&	9.33 	&	 0.4	&	 9.23 	 &	 0.3	 &	 9.13 	\\
F	&	18.9984 	&	4.56 	&		&	4.56 	&		 &	4.56 	 &		 &	 4.56 	\\
Ne	&	20.1797 	&	8.08 	&	0.1	&	8.18 	&	 0.1	&	 8.18 	 &	 0.1	 &	 8.18 	\\
Na	&	22.9898 	&	6.33 	&		&	6.33 	&		 &	6.33 	 &		 &	 6.33 	\\
Mg	&	24.3050 	&	7.58 	&	0.1	&	7.68 	&	 0.1	&	 7.68 	 &	 0.1	 &	 7.68 	\\
Al	&	26.9815 	&	6.47 	&		&	6.47 	&		 &	6.47 	 &		 &	 6.47 	\\
Si	&	28.0855 	&	7.55 	&	0.1	&	7.65 	&	 0.1	&	 7.65 	 &	 0.1	 &	 7.65 	\\
P	&	30.9738 	&	5.45 	&		&	5.45 	&		 &	5.45 	 &		 &	 5.45 	\\
S	&	32.0650 	&	7.33 	&	0.1	&	7.43 	&	 0.1	&	 7.43 	 &	 0.1	 &	 7.43 	\\
Cl	&	35.4530 	&	5.50 	&		&	5.50 	&		 &	5.50 	 &		 &	 5.50 	\\
Ar	&	39.9480 	&	6.40 	&		&	6.40 	&		 &	6.40 	 &		 &	 6.40 	\\
K	&	39.0983 	&	5.12 	&		&	5.12 	&		 &	5.12 	 &		 &	 5.12 	\\
Ca	&	40.0780 	&	6.36 	&	0.1	&	6.46 	&	 0.1	&	 6.46 	 &	 0.1	 &	 6.46 	\\
Sc	&	44.9559 	&	3.17 	&		&	3.17 	&		 &	3.17 	 &		 &	 3.17 	\\
Ti	&	47.8670 	&	5.02 	&	0.1	&	5.12 	&	 0.1	&	 5.12 	 &	 0.1	 &	 5.12 	\\
V	&	50.9415 	&	4.00 	&		&	4.00 	&		 &	4.00 	 &		 &	 4.00 	\\
Cr	&	51.9961 	&	5.67 	&		&	5.67 	&		 &	5.67 	 &		 &	 5.67 	\\
Mn	&	54.9380 	&	5.39 	&		&	5.39 	&		 &	5.39 	 &		 &	 5.39 	\\
Fe	&	55.8450 	&	7.50 	&		&	7.50 	&		 &	7.50 	 &		 &	 7.50 	\\
Co	&	58.9332 	&	4.92 	&		&	4.92 	&		 &	4.92 	 &		 &	 4.92 	\\
Ni	&	58.6934 	&	6.25 	&		&	6.25 	&		 &	6.25 	 &		 &	 6.25 	\\
\enddata
\end{deluxetable}


\clearpage
\begin{deluxetable}{ccc|cccc|ccccccccccc}
\tabletypesize{\scriptsize}
\tablecaption{Comparison Between Metal Mixtures with Similar C and O Abundances, but Different $\alpha$-enhancement Factors \label{tab:comp13}}
\tablehead{
Ele. &  Weight  & $\log$ $N_\odot$ &   Enhancement  &    $\log$ $N$      & N.F. & M.F. &	 Enhancement  &    $\log$ $N$      & N.F. & M.F.
}
\startdata
 	C	  	&		12.0107 	  	&		 8.52 	  	&		0.00   	&		 8.52   	&	 	 0.136327 	 	&		0.097151 	 	&		0.00   	 &		 8.52 	  	&	 	 0.127894 	  	 &		 0.088943   	\\
	N		&		14.0067 		&		7.92 		&		...		 &		 7.92 		 &		0.034244 		&		0.028458 		&	...			 &		 7.92 		&		 0.032126 		&		0.026054 		 \\
	O	  	&		15.9994 	  	&	 8.83   	&		0.40  	&		 9.23 	 	&	 0.699169  	&	0.663698   	&	0.40   	&	9.23 	 	&	 0.655920   	 &	 0.607640 	 	\\
	F		&		18.9984 		&		4.56 		&		...		 &		 4.56 		 &		0.000015 		&		0.000017 		&	...			 &		 4.56 		&		 0.000014 		&		0.000015 		 \\
	Ne	  	&	20.1797 	  	&		 8.08 	 	&		0.10 	  	&	 8.18 	  	&		 0.062313   	&	0.074605 	  	&		0.30   	&	 8.38 	  	 &	 0.092651 	 	 &	 0.108257 	 	\\
	Na		&		22.9898 		&		6.33 		&		...		 &		 6.33 		 &		0.000880 		&		0.001201 		&	...			 &		 6.33 		&		 0.000826 		&		0.001099 		 \\
	Mg  	&	24.3050 	 	&	 7.58 	  	&		0.10 	  	&	 7.68   	&		 0.019705   	&	0.028416 	 	&	0.30 	&	7.88 	 	 &		 0.029299 	  	 &	 0.041232 	  	\\
	Al		&		26.9815 		&		6.47 		&		...		 &		 6.47 		 &		0.001215 		&		0.001945 		&	...			 &		 6.47 		&		 0.001140 		&		0.001781 		 \\
	Si	 	&	28.0855 	 &	 7.55  	&		0.10 	 &	7.65 	 	&	 0.018390 	 	&	0.030644 	 	&	0.30  	&	7.85 	  	&	 	 0.027343 		 & 	 0.044466 	 	\\
	P		&		30.9738 		&		5.45 		&		..		 &		 5.45 		 &		0.000116 		&		0.000213 		&	...			 &		 5.45 		&		 0.000109 		&		0.000195 		 \\
	S	  	&	 	32.0650 	  	&	 	 7.33 	 	&	 	0.10 	 	 & 	7.43  	&	 	 0.011081   	&	 	0.021078 	  	&	 	0.30   	 &	 	 7.63 	  	&	 	 0.016476  	 &	 	 0.030590 	  	\\
	Cl		&		35.4530 		&		5.50 		&		...		 &		 5.50 		 &		0.000130 		&		0.000274 		&		...		 &		 5.50 		&		 0.000122 		&		0.000251 		 \\
	Ar		&		39.9480 		&		6.40 		&		...		 &		 6.40 		 &		0.001034 		&		0.002451 		&		...		 &		 6.40 		&		 0.000970 		&		0.002244 		 \\
	K		&		39.0983 		&		5.12 		&		...		 &		 5.12 		 &		0.000054 		&		0.000126 		&	...			 &		 5.12 		&		 0.000051 		&		0.000115 		 \\
 	Ca	 	& 	40.0780 	 	& 	 6.36 	  	&	  	0.10 	 &	 	 6.46 	 	&	  	 0.001187 	 &	 	0.002824 	 	&	 	0.30 	  	 &		 6.66 	  	&	  	 0.001765 	  	 &	 	 0.004097 	 	\\
	Sc		&		44.9559 		&		3.17 		&		...		 &		 3.17 		 &		0.000001 		&		0.000002 		&	...			 &		 3.17 		&		 0.000001 		&		0.000001 		 \\
 	Ti	  	& 	47.8670 	  	& 	 5.02 	&	 	0.10  	&	 	5.12 	 	 &	 	 0.000054 	 	&	 	0.000154 	  	& 	0.30 	 	&	 	 5.32  &	   0.000081 	  &  0.000224   	\\
	V		&		50.9415 		&		4.00 		&		...		 &		 4.00 		 &		0.000004 		&		0.000012 		&		...		 &		 4.00 		&		 0.000004 		&		0.000011 		 \\
	Cr		&		51.9961 		&		5.67 		&	...			 &		 5.67 		 &		0.000193 		&		0.000594 		&	...			 &		 5.67 		&		 0.000181 		&		0.000544 		 \\
	Mn		&		54.9380 		&		5.39 		&		...		 &		 5.39 		 &		0.000101 		&		0.000329 		&	...			 &		 5.39 		&		 0.000095 		&		0.000302 		 \\
	Fe		&		55.8450 		&		7.50 		&	...			 &		 7.50 		 &		0.013019 		&		0.043139 		&	...			 &		 7.50 		&		 0.012214 		&		0.039495 		 \\
	Co		&		58.9332 		&		4.92 		&	...			 &		 4.92 		 &		0.000034 		&		0.000120 		&	...			 &		 4.92 		&		 0.000032 		&		0.000110 		 \\
	Ni		&		58.6934 		&		6.25 		&	...			 &		 6.25 		 &		0.000732 		&		0.002550 		&	...			 &		 6.25 		&		 0.000687 		&		0.002334 		 \\
\enddata
\tablenotetext{}{The abbreviation N.F. represents the number fraction of each metal element, and M.F. means the mass fraction of each element.}
\end{deluxetable}

\clearpage
\begin{table}
\centering
\caption{ Metallicity Ralation}
\label{tab:compZ}
\begin{tabular}{cccc}
\hline\hline                 

 [Fe/H]   & $Z$                       & $Z$                       \\
     ~           &  [C/{\rm{Fe}}] = 0.0 &  [C/{\rm{Fe}}] = 0.0 \\
                &  [O/{\rm{Fe}}] = 0.4 &  [O/{\rm{Fe}}] = 0.4 \\
                                & $[\alpha /{\rm{Fe}}]$ = 0.1 & $[\alpha /{\rm{Fe}}]$ = 0.3 \\
\noalign{\smallskip}\hline\noalign{\smallskip}
-0.90 	&	0.0038 	&	0.0041 	\\
-0.95 	&	0.0034 	&	0.0037 	\\
-1.00 	&	0.0030 	&	0.0033 	\\
-1.10 	&	0.0024 	&	0.0026 	\\
-1.20 	&	0.0019 	&	0.0021 	\\
-1.30 	&	0.0015 	&	0.0017 	\\
-1.40 	&	0.0012 	&	0.0013 	\\
-1.50 	&	0.0010 	&	0.0011 	\\
    \hline                           
       \end{tabular}
       \end{table}

\begin{deluxetable}{cccccccccccccc}
\tablecaption{Observation Constraints and Modeling Results with Different $\alpha$-enhancement \label{tab:compresult}}
\tablehead{
ID & T$_{\rm{eff}}$ & log $g$  & [Fe/H]   & Age (Gyr)   & Mass ($M_\odot$)   & Age (Gyr)   & Mass ($M_\odot$)  \\
    &(K) & (dex) & (dex) & $\alpha$=0.1 & & $\alpha$=0.3 & &
}
\startdata
 G119-64	&	6333 $\pm$ 35 	&	4.14 $\pm$ 0.06 	&	-1.50 $\pm$ 0.03 	&	 $	 8.40 	 ~_{-~	 0.47	 }^{+~	0.28 	}	 $	 &	$	 0.87 	~_{-~	 0.00	 }^{+~	0.01 	 }	$	 &	 $	 8.04 	 ~_{-~	 0.29	 }^{+~	0.19 	 }	$	&	 $	 0.89 	 ~_{-~	0.00	 }^{+~	 0.01 	}	$	\\
HD 59392	&	6137 $\pm$ 35  	&	3.88 $\pm$ 0.06 	&	-1.62 $\pm$ 0.03 	&	$	 8.86 	 ~_{-~	 0.93	 }^{+~	 1.40 	}	$	&	 $	 0.87 	 ~_{-~	 0.04	 }^{+~	0.03 	 }	$	 &	$	 9.34 	 ~_{-~	 1.47	}^{+~	 0.38 	 }	$	&	 $	 0.87 	~_{-~	0.02	 }^{+~	 0.04 	}	 $	\\
HD 219617	&	5983 $\pm$ 35 	&	4.28 $\pm$ 0.06 	&	-1.46 $\pm$ 0.03 	&	 $	 17.46 	 ~_{-~	 0.69	 }^{+~	0.56 	}	$	 &	 $	 0.70 	~_{-~	 0.00	 }^{+~	0.01 	 }	$	 &	 $	 16.14 	 ~_{-~	 0.78	 }^{+~	0.81 	 }	$	&	 $	 0.72 	 ~_{-~	0.00	 }^{+~	 0.01 	}	$	\\
\enddata
\end{deluxetable}

\allauthors


\begin{thebibliography}{99}



\bibitem[\protect\citeauthoryear{Adibekyan}{2012}]{Adibekyan2012} Adibekyan, V. Zh., Sousa, S. G. , Santos, N. C., et al. 2012, A\&A, 545, A32

\bibitem[\protect\citeauthoryear{Adibekyan}{2013}]{Adibekyan2013} Adibekyan, V. Zh., Figueira, P., Santos, N. C., et~al. 2013, A\&A, 554, A44


\bibitem[\protect\citeauthoryear{Alexander \& Ferguson}{1994}]{AF94} Alexander, D.R \& Ferguson, J.W. 1994, ApJ, 437, 879


\bibitem[\protect\citeauthoryear{Barnes}{2007}]{Barnes2007} Barnes, S. A. 2007, ApJ, 669, 1167

\bibitem[\protect\citeauthoryear{Basu et~al.}{2010}]{Basu2010} Basu, S., Chaplin, W. J., \& Elsworth, Y., 2010, ApJ, 710, 1596


\bibitem[\protect\citeauthoryear{Bennett et~al.}{2013}]{Bennett2013} Bennett, C. L., Larson, D., Weiland, J. L., et~al. 2013, ApJS, 208, 20


\bibitem[\protect\citeauthoryear{Bensby et~al.}{2003}]{Bensby2003} Bensby, T., Feltzing, S., Lundstrom, I., 2003, A\&A, 410, 527

\bibitem[\protect\citeauthoryear{Bonavita \& Desidera}{2007}]{BD2007} Bonavita, M., \& Desidera, S., 2007, A\&A, 468, 721


\bibitem[\protect\citeauthoryear{Bovy et~al.}{2012}]{Bovy2012} Bovy, J., Rix, H.-W., \& Hogg, D. W. 2012, ApJ, 751, 131

\bibitem[\protect\citeauthoryear{Bovy et~al.}{2016}]{Bovy2016} Bovy, J., Rix, H.-W., Schlafly, E. F., et~al. 2016, ApJ, 823, 30


\bibitem[\protect\citeauthoryear{Carney et~al.}{2001}]{Carney2001} Carney, B.W., Latham, D.W., Laird, J. B., Grant, C. E., \& Morse, J. A. 2001, AJ, 122, 3419


\bibitem[\protect\citeauthoryear{Demarque et~al.}{2004}]{Demarque2004} Demarque, P., Woo, J.-H., Kim, Y.-C., \& Yi, S. K. 2004, ApJS, 155, 667

\bibitem[\protect\citeauthoryear{Demarque et~al.}{2008}]{Demarque2008} Demarque, P., Guenther, D. B., Li, L. H., Mazumdar, A., Straka, C. W., 2008, Ap\&SS, 316, 31D


\bibitem[\protect\citeauthoryear{Deng et~al.}{2012}]{Deng2012} Deng L.-C., Newberg, H. J., Liu, C., et~al. 2012, RAA, 12, 735

\bibitem[\protect\citeauthoryear{Dotter et~al.}{2008}]{Dotter2008} Dotter, A., Chaboyer, B,, Jevremovi\'c, D., Kostov, V., Baron, E., Ferguson, J. W., 2008, ApJS, 178, 89

\bibitem[\protect\citeauthoryear{Eggen et~al.}{1962}]{Eggen1962} Eggen, O. J., Lynden-Bell, D., \& Sandage, A. R., 1962, ApJ, 136, 748

\bibitem[\protect\citeauthoryear{Ferguson et al.}{2005}]{Ferguson2005}Ferguson, J. W., Alexander, D. R., Allard, F., et al., 2005, ApJ, 623, 585


\bibitem[\protect\citeauthoryear{Fuhrmann}{2008}]{Fuhrmann2008} Fuhrmann, K., 2008, MNRAS, 384, 173

\bibitem[\protect\citeauthoryear{Ge et~al.}{2015}]{Ge2015} Ge, Z. S., Bi, S. L., Li, T. D., et~al. 2015, MNRAS, 447, 680


\bibitem[\protect\citeauthoryear{Gilmore \& Reid}{1983}]{GR1983} Gilmore, G., \& Reid, N., 1983, MNRAS, 202, 1025

\bibitem[\protect\citeauthoryear{Girardi et~al.}{2000}]{Girardi2000} Girardi, L., Bressan, A., Bertelli, G., et~al. 2000, A\&AS, 141, 371

\bibitem[\protect\citeauthoryear{Grevesse \& Noels}{1993}]{GN93}    Grevesse, N., \& Noels, A. 1993, in Origin and Evolution of the Elements, ed. N. Prantzos, E. Vangioni-Flam, \& M. Casse′ (Cambridge: Cambridge Univ. Press), 15


\bibitem[\protect\citeauthoryear{Grevesse \& Sauval}{1998}]{GS98} Grevesse, N., \& Sauval, A. J. 1998, Space Sci. Rev., 85, 161


\bibitem[\protect\citeauthoryear{YREC, Guenther et~al.}{1992}]{Guenther1992} Guenther, D. B., Demarque, P., Kim, Y.-C., et~al. 1992, ApJ, 387, 372


\bibitem[\protect\citeauthoryear{Hawkins et~al.}{2014}]{Hawkins2014} Hawkins, K., Jofr\'e, P., Gilmore, G., et~al. 2014, MNRAS, 445, 2575

\bibitem[\protect\citeauthoryear{Haywood et~al.}{2013}]{Haywood2013} Haywood, M., Di Matteo, P., Lehnert, M. D., et~al. 2013, A\&A, 560, 109


\bibitem[\protect\citeauthoryear{Ibata, Gilmore \& Irwin}{1994}]{Ibata1994} Ibata R. A., Gilmore G., Irwin M. J., 1994, Nature, 370, 194


\bibitem[\protect\citeauthoryear{Jofr\'e \& Weiss}{2011}]{JW2011} Jofr\'e, P., Weiss, A., 2011, A\&A, 533, A59

\bibitem[\protect\citeauthoryear{Juri\'c et~al.}{2008}]{J2008}  Juri\'c, M., Ivezi\'c, \v Z., Brooks, A., et al. 2008, ApJ, 673, 864

\bibitem[\protect\citeauthoryear{Kallinger et~al.}{2010}]{Kallinger2010} Kallinger, T., Mosser, B., Hekker, S., et~al. 2010, A\&A, 522, A1

\bibitem[\protect\citeauthoryear{Kaufer et~al.}{1999}]{Kaufer1999} Kaufer, A., Stahl, O., Tubbesing, K., et~al. 1999, The Messenger, 95, 8


\bibitem[\protect\citeauthoryear{Kim et~al.}{2002}]{Kim2002} Kim, Y. -C., Demarque, P., Yi, S.K. \& Alexander, D.R. 2002, ApJS, 143, 499

\bibitem[\protect\citeauthoryear{Kobayashi et~al.}{2006}]{Kobayashi2006} Kobayashi, C., Umeda, H., Nomoto, K., Tominaga, N., \& Ohkubo, T. 2006, ApJ, 653, 1145


\bibitem[\protect\citeauthoryear{Lindegren et~al.}{2010}]{Lindegren2010} Lindegren, L. 2010, IAU Symp. 261, Relativity in Fundamental Astronomy: Dynamics, Reference Frames, and Data Analysis, eds. S. A. Klioner, P.K. Seidelmann, \& M.H. Soffel (Cambridge: Cambridge Univ. Press), 296


\bibitem[\protect\citeauthoryear{Liu et~al.}{2014}]{Liu2014} Liu, X. W., Yuan, H. B., Huo, Z. Y., et al. 2014, in Proc. IAU Symp. 298,
Setting the Scene for Gaia and LAMOST, ed. S. Feltzing et al. (Cambridge:
Cambridge Univ. Press), 310

\bibitem[\protect\citeauthoryear{Mayor et~al.}{2003}]{Mayor2003} Mayor, M., Pepe, F., Queloz, D., et al. 2003, The Messenger , 114, 20

\bibitem[\protect\citeauthoryear{Nissen \& Schuster}{2010}]{NS10} Nissen, P. E., \& Schuster, W. J. 2010, A\&A, 511, L10

\bibitem[\protect\citeauthoryear{Nissen \& Schuster}{2011}]{NS11} Nissen, P. E., \& Schuster, W. J. 2011, A\&A, 530, A15

\bibitem[\protect\citeauthoryear{Nissen \& Schuster}{2012}]{NS12} Nissen, P. E., \& Schuster, W. J. 2012, A\&A, 543, A28

\bibitem[\protect\citeauthoryear{Nissen et~al.}{2014}]{Nissen14} Nissen, P. E., Chen, Y. Q., Garigi, L., et~al. 2014, A\&A, 568, 25

\bibitem[\protect\citeauthoryear{Nordstr\"om et~al.}{2004}]{Nordstrom2004} Nordstr\"om, B., Mayor, M., Andersen, J., et al. 2004, A\&A, 418, 989


\bibitem[\protect\citeauthoryear{Perryman et~al.}{2001}]{Perryman2001} Perryman, M. A. C., de Boer, K. S., Gilmore, G., et~al. 2001, A\&A, 369, 339

\bibitem[\protect\citeauthoryear{Ram\'irez et~al.}{2012}]{Ramirez12} Ram\'irez, I., Mel\'endez J., \& Chanam\'e, J., 2012, ApJ., 757, 164

\bibitem[\protect\citeauthoryear{Ram\'irez et~al.}{2013}]{Ramirez13} Ram\'irez, I., Allende P., C., \& Lambert, D. L., 2013, ApJ., 764, 78

\bibitem[\protect\citeauthoryear{Rocha-Pinto \& Maciel}{1998}]{RP1998} Rocha-Pinto, H. J., \& Maciel, W. J. 1998, MNRAS, 298, 332

\bibitem[\protect\citeauthoryear{Rogers \& Nayfonov}{2002}]{Rogers2002} Rogers, F. J., Nayfonov, A., 2002, ApJ, 576, 1064

\bibitem[\protect\citeauthoryear{Salaris et~al.}{2006}]{Salaris2006} Salaris, M., Weiss, A., Ferguson, J. W., et~al. 2006, ApJ, 645, 1131

\bibitem[\protect\citeauthoryear{Salasnich et~al.}{2000}]{Salasnich2000} Salasnich, B., Girardi, L., Weiss, A., et al., 2000, A\&A, 361, 1023

\bibitem[\protect\citeauthoryear{Schr\"oder et~al.}{2013}]{Schroder2013} Schr\"oder, K.-P., Mittag, M., Hempelmann, A., et~al. 2013, A\&A, 554, 50

\bibitem[\protect\citeauthoryear{Schuster et~al.}{2012}]{Schuster2012} Schuster, W. J., Moreno, E., Nissen, P. E., et al., 2012, A\&A, 538, A21

\bibitem[\protect\citeauthoryear{Searle \& Zinn}{1978}]{SearleZinn1978} Searle, L., \& Zinn, R., 1978, ApJ, 225, 357

\bibitem[\protect\citeauthoryear{Sitnova et~al.}{2015}]{Sitnova2015}   Sitnova, T., Zhao, G., Mashonkina, L., et~al. 2015, ApJ, 808, 148

\bibitem[\protect\citeauthoryear{Spergel et~al.}{2007}]{Spergel2007}Spergel, D. N., Bean, R., Dor′e, O., et~al. 2007, ApJS, 170, 377

\bibitem[\protect\citeauthoryear{Thoul et~al.}{1994}]{Thoul1994} Thoul, A. A., Bahcall, J. N., \& Loeb, A. 1994, ApJ, 421, 828

\bibitem[\protect\citeauthoryear{Valenti \& Fischer}{2005}]{VF2005} Valenti, J. A., \& Fischer, D. A. 2005, ApJS, 159, 141

  \bibitem[\protect\citeauthoryear{VandenBerg et~al.}{2000}]{VandenBerg2000} VandenBerg, D. A., Swenson, F. J., Rogers, F. J., Iglesias, C. A., \& Alexander, D. R. 2000, ApJ, 532, 430


\bibitem[\protect\citeauthoryear{VandenBerg et~al.}{2014}]{VandenBerg2014} VandenBerg, D. A., Bond, H. E., Nelan, E. P., et~al. 2014, ApJ, 792, 110

\bibitem[\protect\citeauthoryear{Ventura et~al.}{2009}]{Ventura2009} Ventura, P., Caloi, V.,   D’Antona, F.,   Ferguson, J. W., Milone, A., et~al. 2009, MNRAS, 399, 934

\bibitem[\protect\citeauthoryear{Yang \& Bi}{2007}]{YangBi2007} Yang, W. M., \& Bi, S. L.,  2007, ApJL, 658, L67


\bibitem[\protect\citeauthoryear{Yanny et~al.}{2009}]{Yanny2009} Yanny, B., Newberg, H. J., Johnson, J. A., et~al. 2009, ApJ, 700, 1282

\bibitem[\protect\citeauthoryear{Yi et~al.}{2001}]{Yi2001} Yi, S., Demarque, P., Kim, Y. -C., Lee, Y.-W., Ree, C.H., Lejeune, Th. \& Barnes, S. 2001, ApJS, 136, 417

\bibitem[\protect\citeauthoryear{Yi et~al.}{2003}]{Yi2003} Yi, S. K., Kim, Y.-C., \& Demarque, P. 2003, ApJS, 144, 259

 \bibitem[\protect\citeauthoryear{Zolotov et~al.}{2009}]{Zolotov2009} Zolotov, A., Willman, B., Brooks, A. M., et~al. 2009, ApJ, 702, 1058
 \bibitem[\protect\citeauthoryear{Zolotov et~al.}{2010}]{Zolotov2010} Zolotov, A., Willman, B., Brooks, A. M., et~al. 2010, ApJ, 721, 738

\bibitem[\protect\citeauthoryear{Zhao et~al.}{2012}]{Zhao2012} Zhao, G., Zhao, Y. H., Chu, Y. Q., et~al. 2012, RAA, 12, 723

\end{thebibliography}
\end{document}